\documentclass[manuscript,screen]{acmart}
\AtBeginDocument{%
  \providecommand\BibTeX{{%
    \normalfont B\kern-0.5em{\scshape i\kern-0.25em b}\kern-0.8em\TeX}}}

\setcopyright{acmcopyright}
\copyrightyear{2018}
\acmYear{2018}
\acmDOI{XXXXXXX.XXXXXXX}

\acmConference[Conference acronym 'XX]{Make sure to enter the correct
  conference title from your rights confirmation emai}{June 03--05,
  2018}{Woodstock, NY}
\acmBooktitle{Woodstock '18: ACM Symposium on Neural Gaze Detection,
 June 03--05, 2018, Woodstock, NY} 
\acmPrice{15.00}
\acmISBN{978-1-4503-XXXX-X/18/06}
\usepackage{tabularx} 
\usepackage{booktabs} 
\usepackage{svg}
\usepackage{hyperref} 
\usepackage{cleveref}
\usepackage{listings}
\usepackage{xcolor}

\begin{document}

\title{Set Visualizations for Comparing and Evaluating Machine Learning Models}

\author{Liudas Panavas}
\email{panavas.l@northeastern.edu}
\orcid{0000-0003-0428-5579}
\author{Tarik Crnovrsanin}
\email{t.crnovrsanin@northeastern.edu}
\orcid{0000-0002-4397-5532}
\author{Racquel Fygenson}
\email{fygenson.r@northeastern.edu}
\orcid{0000-0002-0705-9000}
\affiliation{%
  \institution{Northeastern University}
  \streetaddress{440 Huntington Avenue}
  \city{Boston}
  \state{Massachusetts}
  \country{USA}
  \postcode{02115}
}

\author{Eamon Conway}
\email{e.conway@northeastern.edu}
\affiliation{%
    \institution{Kostas Research Institute at Northeastern University}
    \streetaddress{141 South Bedford Street}
    \city{Burlington}
    \state{Massachusetts}
    \country{USA}
    \postcode{01803}
}

\author{Derek M.\ Millard}
\email{derek.m.millard.civ@army.mil}
\author{Norbou Buchler}
\email{norbou.buchler.civ@army.mil}
\affiliation{%
    \institution{U.S.\ Army Combat Capabilities Development Command Analysis Center}
    \streetaddress{}
    \country{USA}
}

\author{Cody Dunne}
\email{c.dunne@northeastern.edu}
\orcid{0000-0002-1609-9776}
\affiliation{%
  \institution{Northeastern University}
  \streetaddress{440 Huntington Avenue}
  \city{Boston}
  \state{Massachusetts}
  \country{USA}
  \postcode{02115}
}

\renewcommand{\shortauthors}{Panavas, et al.}

\begin{abstract}
  Machine learning practitioners often need to compare multiple models to select the best one for their application. However, current methods of comparing models fall short because they rely on aggregate metrics that can be difficult to interpret or do not provide enough information to understand the differences between models. To better support the comparison of models, we propose set visualizations of model outputs to enable easier model-to-model comparison. We outline the requirements for using sets to compare machine learning models and demonstrate how this approach can be applied to various machine learning tasks. We also introduce SetMLVis, an interactive system that utilizes set visualizations to compare object detection models. Our evaluation shows that SetMLVis outperforms traditional visualization techniques in terms of task completion and reduces cognitive workload for users. Supplemental materials can be found at \url{https://osf.io/afksu/?view_only=bb7f259426ad425f81d0518a38c597be}.
\end{abstract}

\ccsdesc[500]{Computing methodologies~Object detection}
\ccsdesc[500]{Human-centered computing~Visualization toolkits}
\ccsdesc[500]{Human-centered computing~Visualization theory, concepts and paradigms}
\ccsdesc[500]{Human-centered computing~Empirical studies in visualization}
\keywords{Set visualization, Machine learning model comparison, Object detection evaluation, Visual analytics, Model interpretability}


\received{24 January 2024}
\received[revised]{12 March 2009}
\received[accepted]{5 June 2009}

\maketitle

\section*{Bottom Line Up Front}

\begin{enumerate}
    \item \textbf{Implications for researchers:} We introduce a paradigm shift in model comparison using set theory and visualization, enabling direct comparison of model predictions. This approach enhances interpretability and provides a foundation for new exploratory data analysis techniques, allowing researchers to uncover nuanced insights and innovate in model evaluation.
    \item \textbf{Implications for ML practitioners:} Our paper provides practical code examples demonstrating how to apply set theory and visualization to a variety of machine learning model types. Additionally, our interactive visualization tool, SetMLVis, is specifically designed for object detection models, helping practitioners quickly identify strengths and weaknesses, streamline evaluation, and make better-informed decisions. (Software links: \href{https://pypi.org/project/setmlvis/}{PyPI}, \href{https://github.com/VisDunneRight/setmlvis}{GitHub})
\end{enumerate}

\begin{figure}[!htb] 
    \centering
    \includegraphics[width=\linewidth]
    {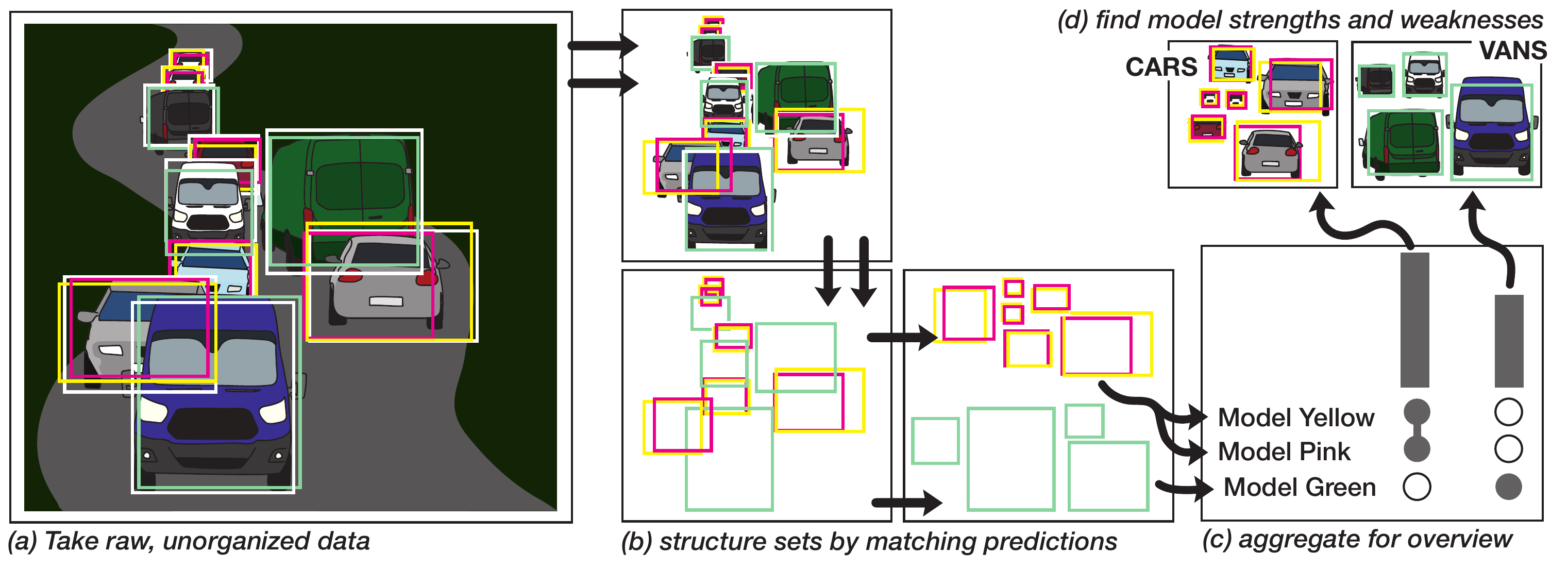}
  \caption{When comparing object detection models, practitioners look at bounding box overlays on images (panel (a)). These visualizations can quickly become convoluted and make extracting meaningful comparisons difficult. By matching predictions from the different models and placing them into sets (panel (b)), we can create visualizations that make it easier to compare the models. Using an Upset style visualizations (panels (c)) the predictions can be analyzed through meaningful subsets of the data. For instance, examining the predictions in the two bars shows  Models Yellow and Pink excel in detecting cars, while Model Green is adept at identifying vans (panel (d)). This innovative use of set visualizations provides clear, actionable insights when comparing model performance.}
  \Description{This figure compares object detection models using set visualizations. It begins with bounding box overlays from three models on an image in panel (a). Panel (b) organizes predictions into sets, and panel (c) presents an Upset plot showing aggregated predictions. Panel (d) highlights how specific models excel at detecting different vehicle types, such as cars and vans.}
  \label{fig:teaser}
\end{figure}

\section{Introduction}

To deploy an effective machine learning application, practitioners must select the most suitable model for their specific needs \cite{sacha2016human}. This choice may involve evaluating out-of-the-box solutions from open-source platforms or customizing models through parameter adjustments. Regardless of the development path taken, practitioners are often faced with several models to choose from. To make sense of the models and select the right one, they can perform a comparative analysis to deepen their understanding of each model's capabilities and select the model best suited for their application \cite{borji2015salient}.

\begin{figure}[!htb] 
    \centering
    \includegraphics[width=\linewidth]{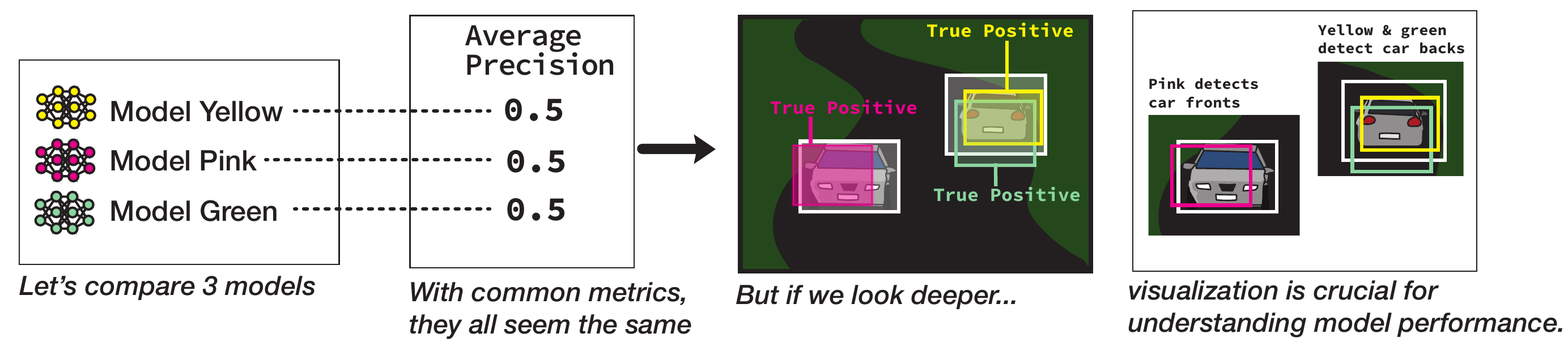}
    \caption{While all three models have the same average precision of 0.5, this metric hides key differences in their predictions. A closer look at the instance-level data shows that Model Pink consistently detects car fronts, whereas Models Yellow and Green focus on car backs. Relying on aggregate metrics alone misses these important distinctions, which become apparent only through detailed analysis of model outputs.}
    \label{fig:motivation}
\end{figure}

Although comparing machine learning models is an essential step, current tools and techniques face challenges. Common metrics like precision and recall can be difficult to interpret, result in similar values, and fail to reveal the underlying causes of performance differences \cite{oksuz2018localization, reinke2021common, peng2021systematic, gortler2022neo}. For instance, a model tasked with traffic surveillance may be suited for detecting fronts of cars, while another excels with backs of cars; however, the aggregate metrics cannot differentiate between these strengths (Figure \ref{fig:motivation} A and B). Examining image-level predictions provides interpretability, but is time-consuming. Datasets often contain thousands of images, each with numerous detections. Additionally, significant overlap in model predictions can make thorough analysis challenging (\cref{fig:teaser}
(a)). To overcome these challenges, new methods and systems need to be developed for model comparisons \cite{oksuz2018localization}.

Current methods for comparing machine learning models often fall short because they do not have ways to directly compare predictions from different models. Typically, each model is first evaluated against the ground truth to generate aggregate metrics, which are then used for comparison \cite{padilla2020survey}. This indirect approach can obscure significant differences between models (\cref{fig:motivation}
front vs. back of cars) \cite{oksuz2018localization, reinke2021common}. A more effective approach \textit{reverses} this process: first, match the predictions of multiple models to directly compare them and then evaluate them against the ground truth. 
This method results in highlighting areas of agreement and divergences between models and can isolate interesting data instances, simplifying the time consuming task of viewing instance level data (Figure \ref{fig:teaser}) \cite{hohman2018visual, moore2023failurenotes}. After establishing these comparative differences, the models can still be assessed against the ground truth. This revised order—model-to-model comparison first, followed by ground truth comparison—not only clarifies the differences between models but also maintains the integrity of evaluating models against the ground truth.

While some machine learning development techniques match model predictions \cite{dietterich2000ensemble, casado2020ensemble} and visual analytics systems allow people to visually directly compare predictions \cite{cabrera2023did}, to the best of our knowledge, there is no formal method to directly compare model predictions for evaluation and comparison. To fill this gap, we propose using sets and set visualization to enable model-to-model comparison. Sets can be created by matching predictions across models (Figure \ref{fig:teaser}). For instance, matching predictions from models A and B yields three sets: unique predictions from Model A, unique predictions from Model B, and their shared predictions. These sets naturally lead to questions such as, “Which predictions is model A making differently than model B?” This can be determined by examining the set of predictions that model A makes, which model B does not. Analyzing this set can provide insights into each model's strengths and weaknesses (Figure \ref{fig:teaser} Answer: Green model - vans). Furthermore, transforming model outputs into set type data allows for the use of set visualizations, such as UpSet \cite{lex2014upset}, to easily explore the data. These visualizations allow for an examination of the subsets of the data that cause different models to behave differently and reveal problematic areas or avenues for future model development. 

In this work, we outline the fundamental requirements for utilizing sets to compare machine learning models. We demonstrate the process of transforming model data into set-type data, and show how this can enhance comparisons across four distinct types of machine learning. To fully explore set visualizations for machine learning model comparison, we developed SetMLVis, an interactive system that uses set visualizations to compare object detection models. We demonstrate the efficacy of using set visualizations by comparing our system against a leading open-source software that uses ``traditional visualization''. Our evaluation adopts a mixed-method approach, combining quantitative and qualitative measures. Participants completed predefined tasks with both systems that represent model evaluation and comparison tasks found in the literature. Their performance is measured in terms of accuracy and cognitive workload. We utilize both observational techniques and self-reported data to gain insights into user interaction and mental effort. Our results indicate that SetMLVis consistently outperforms the other system on task completion and reduces users cognitive workload. Based on these findings we make three contributions:

\begin{enumerate}
 \item \textbf{Methodology and application of sets in machine learning model comparison:} We introduce a novel methodology that transforms model predictions into sets, enabling direct comparison of outputs across multiple models. This approach leverages set theory to enhance interpretability and provides clear insights into each model's strengths and weaknesses.
\item \textbf{SetMLVis: Object detection model comparison using set visualizations} We develop SetMLVis, an open-source tool that applies set visualizations innovatively for evaluating object detection models, integrating a Python API with an interactive UI.
\item \textbf{Evaluation of set methodology for algorithmic model comparison:} We run a mixed-methods evaluation comparing SetMLVis to “traditional” visualization techniques indicating that set visualizations have practical value for ML model evaluation. In particular, we find improved user performance on ML model comparison tasks and reduced cognitive workload.
\end{enumerate}

Supplementary materials are available, including the SetMLVis tool, detailed case studies, and the complete resources necessary to replicate the user study. These materials encompass the dataset, experimental design, and full analysis of the results. The supplementary materials can be accessed at \url{https://osf.io/afksu/?view_only=bb7f259426ad425f81d0518a38c597be}.

\section{Related Work}
\subsection{Machine Learning Model Comparison}

The comparison of machine learning models is pivotal in determining their effectiveness and applicability in real-world scenarios. Practitioners use aggregate metrics on benchmark datasets \cite{everingham2010pascal}, visual breakdowns of errors \cite{davis2006relationship}, and view instance level data \cite{suh2022visualization} to compare the models. While each of these strategies offers unique insights into model performance, combining them is essential to achieve a nuanced understanding of how well a model functions across different scenarios \cite{sun2020dfseer, nushi2018towards, patel2008investigating}.

\begin{figure}[htbp] 
    \centering
    \includegraphics[width=0.9\linewidth]{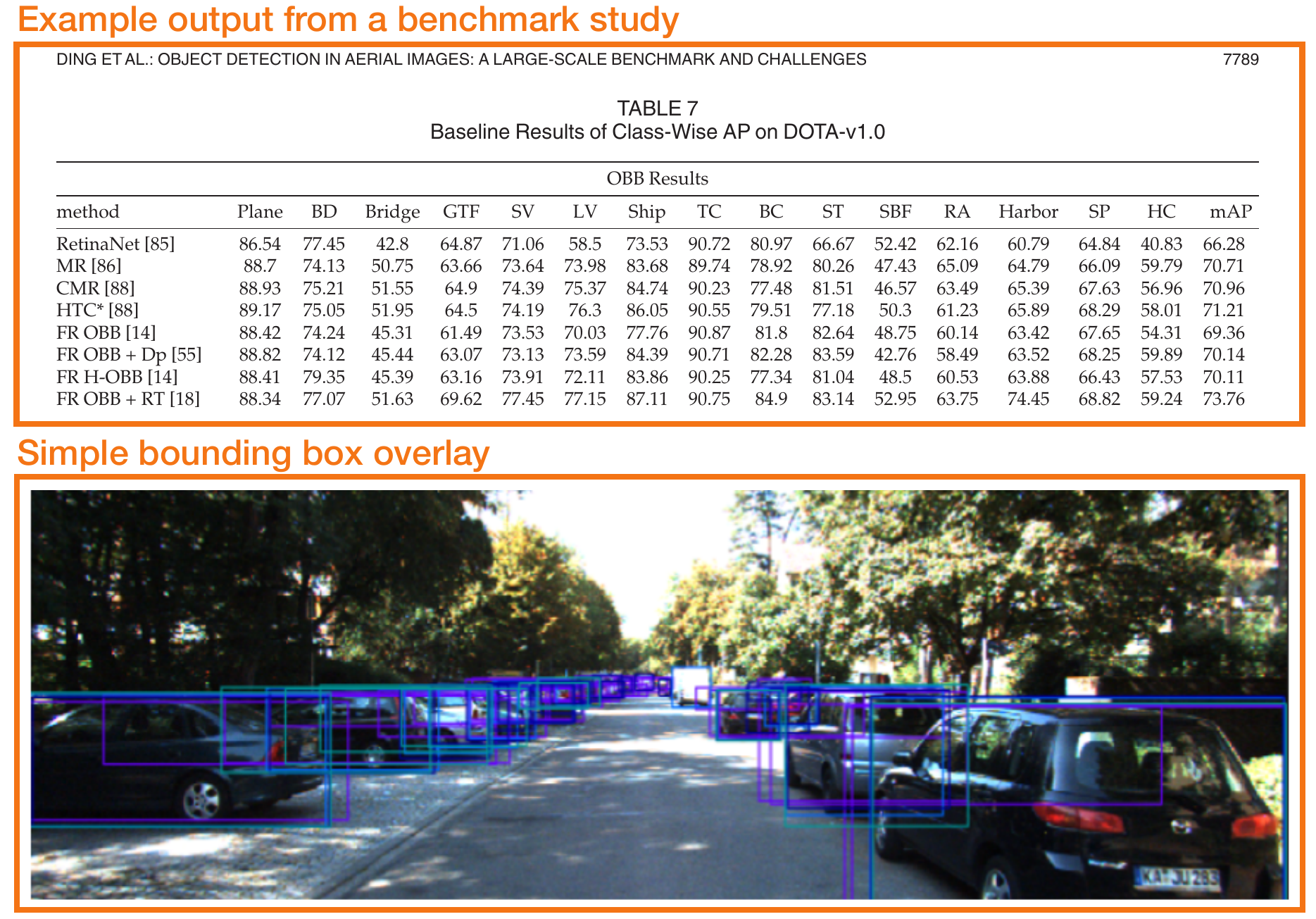}
    \caption{On the top, we see the results of a benchmark study on object detection models where each row represents a model and the columns are precision for a specific class \cite{ding2021object}. While these types of tables are popular, it can be difficult to grasp the differences between models. On the bottom, the challenges from viewing instance-level predictions are apparent. Predictions for cars from Faster R-CNN \cite{ren2015faster}, ResNet \cite{he2016deep}, and DETR \cite{carion2020end} are shown. The dense overlays make it difficult to compare and understand the performance differences between these models.}
    \label{fig:standardEvaluation}
\end{figure}

Aggregate metrics provide a high-level overview of model performance, synthesizing complex data into digestible and comparative numbers that guide practitioners in their evaluation efforts. Metrics, such as average precision (AP), are used to evaluate and compare models in a variety of ML applications (object detection \cite{everingham2010pascal, dai2016r, lin2014microsoft, padilla2020survey}, information retrieval systems \cite{sanderson2005information}, and binary classification \cite{vujovic2021classification}). 
They provide a usable first view and valuable quality check but often fail to differentiate models when performance differences are small. They also lack context by not showing instance-level data, making it difficult to understand the trade-offs and overall effectiveness of the models being compared \cref{fig:standardEvaluation} top) \cite{oksuz2018localization}. For example, multiple models could make entirely different predictions but still receive the same aggregate metric (Figure \ref{fig:motivation}). Therefore, practitioners rely on more granular insights by viewing instance level data. 

Instance-level prediction viewing offers direct insights into the decision-making process of models, facilitating deeper analysis and more effective troubleshooting \cite{nushi2018towards, moore2023failurenotes}.
By viewing the model predictions on each instance, practitioners can better understand aggregate metrics and how the model will perform on a variety of use cases \cite{cabrera2023did, alexander2015task, suh2022visualization}. Methods such as visualizing bounding box overlays on images allow for a clear visual of the model outputs. While viewing these instance predictions is imperative, it can become very time-consuming as practitioners must parse through thousands of instances to identify patterns, anomalies, and outliers crucial to refine model accuracy and reliability  (\cref{fig:standardEvaluation} bottom)\cite{hopkins2021machine}. 

The shortcomings of each evaluation method underscore a critical point: practitioners need both aggregate metrics for a general overview of performance along with specific predictions to compare models thoroughly \cite{suh2022visualization}. However, most approaches to model comparisons and many visual analytics systems continue to depend primarily on aggregate metrics and their visual representations \cite{ding2021object, padilla2020survey}. Therefore, to overcome the shortcomings of the current evaluation methods, we introduce a new method to compare ML models using sets. This method provides high-level metrics that help compare models and then easily connect to instance-level data by automatically generating schemas that intuitively answer practitioner questions. Our visual analytics tool, SetMLVis, demonstrates our proposed method in a user-friendly interface that allows for a thorough evaluation of object detection models.

\subsection{Visual Analytics Systems for Evaluation \& Comparison}

The explosion of machine learning models and complexity of architectures has required new ways to understand, compare, and ultimately choose the best machine learning model for a given task. Visual analytics systems have helped by explaining how a model comes to its prediction \cite{vondrick2013hoggles} or assessing the quality of the outputs \cite{cabrera2023did}. While these systems have made important strides in model interpretability and evaluation, few have allowed for a thorough comparison of multiple models.

A variety of visual analytics systems have been created to make sense and explore the large complex datasets generated by models. Systems such as Neo \cite{gortler2022neo} and Zeno \cite{cabrera2023zeno} help users explore the model outputs by deconstructing aggregate metrics to show hidden classes or explore data slices. Despite their utility, neither Neo nor Zeno supports instance-level data inspection or facilitates inter-model comparisons. Other visual analytics systems have overcome some of these shortcomings. MLCube \cite{kahng2016visual} and Gestalt \cite{patel2010gestalt}, bridge the gap between high-level metrics and individual data points by isolating interesting subsets of the instance-level data to view. This eases the cognitive load by allowing for targeted data inspection \cite{patel2008investigating}. None of these systems allow users to compare models holistically by allowing for both aggregate metric and instance-level comparisons across multiple models. 

In addition to these visual analytic systems, frameworks have been designed to guide new system development. Notably, Cabrera et al.\ \cite{cabrera2023did} outline a four-stage sensemaking framework—Instances and Outputs, Schemas, Hypothesis, and Assessment - that describes how model developers develop mental models of model behavior. Using this framework, they design a visual analytics system, AIFinnity, to evaluate and compare image captioning models. AIFinnity allows for model comparison and makes it easy to compare at the instance level data. While we use their framework to guide our system development, AIFinnity falls short as it only allows for two models to be compared at a time and does not have overview visuals to guide data exploration.

These existing systems provide a basis for our object detection model comparison system. While Zeno and Neo provide clear breakdowns of performance metrics and help isolate specific data segments, they lack the capability for detailed image-level evaluation \cite{cabrera2023zeno, gortler2022neo}. AIFinnity addresses this by offering image-level analysis but requires extensive manual data exploration to create data schemas \cite{cabrera2023did}. Our system synthesizes these approaches by automating the generation of relevant data slices through model comparisons and enabling instance-level prediction comparisons. This allows for both high-level metric analysis and granular image-level insights filling the gap of a holistic model comparison visual analytics system.


\section{Set Visualization for ML evaluation}

Set analysis is extensively used in various domains to interpret and analyze large data collections effectively. Defined as a collection of items linked by common characteristics, sets enable the exploration of data properties, relationships, and patterns that are often obscured in raw data \cite{lex2014upset, alper2011design}. As a simple example, a dataset of animal images could be labeled by their habitat and animal class. Using sets, a scientist could quickly pull up the set of images labeled 'forest' and 'mammal' and the set of images labeled 'grassland' and 'mammal'. Looking at the images they could see the similarities and differences between the mammals in both types of environments to find interesting patterns. 

\begin{figure}[!h]
    \centering
    \includegraphics[width=0.9\textwidth]{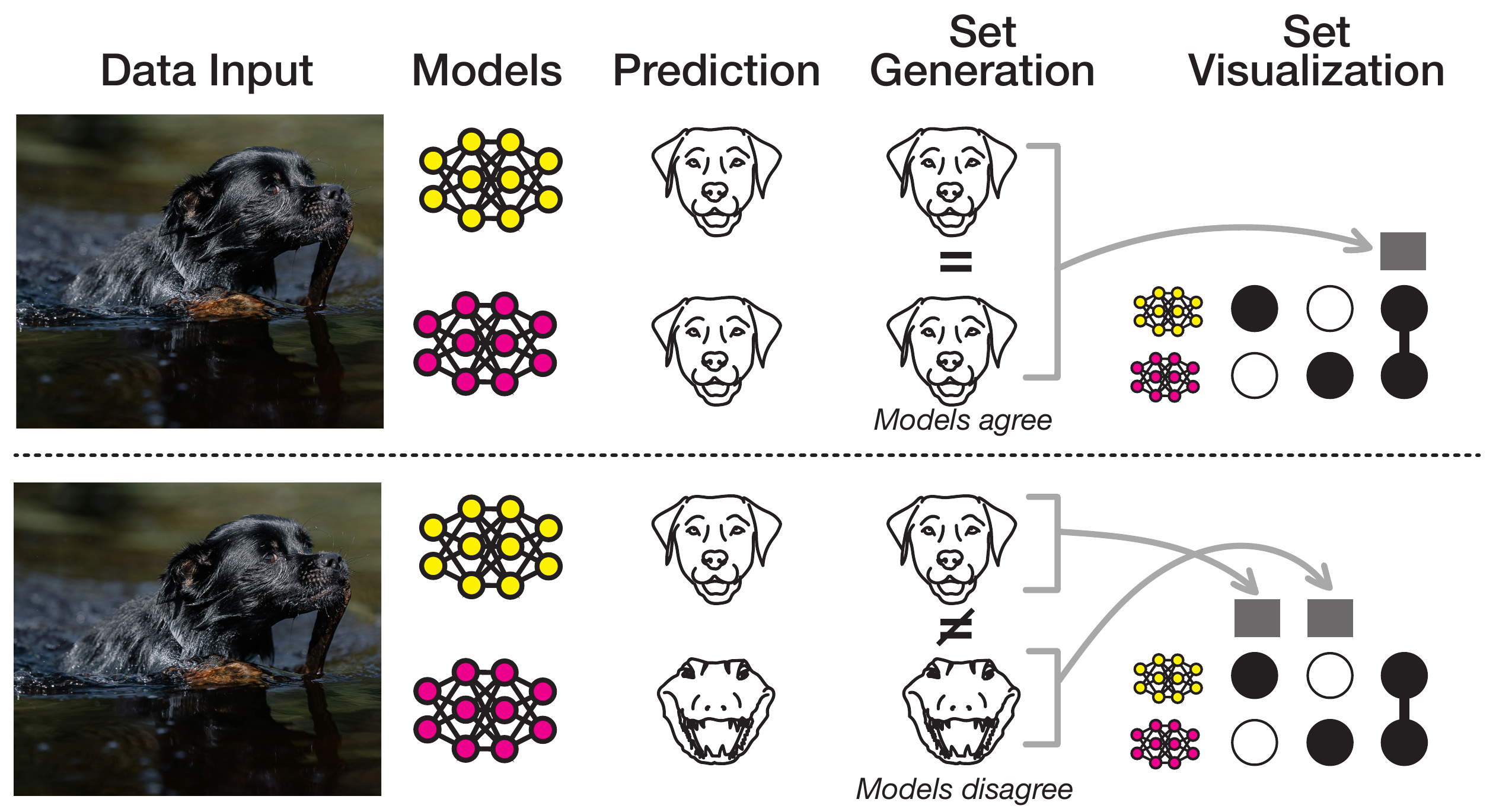}
    \caption{Illustration of set creation from model predictions. Top: The image is classified as a dog by both models, forming a set of model agreement. Bottom: Discrepant predictions from Model A (mammal) and Model B (reptile) lead to the formation of distinct sets for each model's unique prediction.}
    \label{fig:foxExample}
\end{figure}

This process closely parallels model comparison in machine learning. When models are benchmarked against the same dataset, they generate large, complex datasets of predictions. To effectively compare models, practitioners must find which predictions the models agree or disagree on, akin to grouping items in set analysis. These groupings based on agreement or disagreement with other models help reveal each model's strengths and weaknesses—patterns that are not immediately apparent in the raw data.

Let’s see how this works in a similar example to the set analysis described earlier. Instead of an already labeled animal image dataset, two image classification models now label each image as a particular animal class. The scientist first needs to decide which model to choose so they can study they can use good data to perform their animal attribute analysis later. Transforming the data into sets requires a slight alteration of the way sets are commonly thought of. In the original example, an image of a dog would go into the set of mammals. When creating sets from model predictions, if model A and model B both predicted a mammal, it would go into the set of model A and B agreement (\cref{fig:foxExample} (top)). If they predicted different labels (Model A - mammal, Model B - reptile), the image would be added to the sets of unique Model A and Model B predictions (\cref{fig:foxExample} (bottom)).  Later, the model developer can look at the sets of unique predictions and find patterns in model predictions. For instance, Model A predicts mammals better than model B if they are in the water.

Despite these clear parallels, the application of set visualizations in machine learning has not been widely explored, and a systematic methodology for their use is still lacking \cite{agarwal2020set}. 
To fill this gap, we formalize the criteria necessary for machine learning outputs to be transformed into set type data. While not all machine learning outputs are equally suitable for set exploration, if the following four criteria are met the model outputs can be transformed into set data and placed into set visualizations for further explorations. 

\begin{enumerate}
    \item There must be more than one model.
    \item The models must make predictions on the exact same data.
    \item The models need a common output format.
    \item  A well-defined rule or property can determine whether models agree on a prediction or not (set generation rule)
\end{enumerate}

If this criteria is met, model outputs can be transformed into set type data. Transforming the model output data into set type data automatically generates schemas that help in model comparison. These schemas naturally answer a variety of different questions about the models and can be explored through set visualizations.

\subsection{Visualizing Intersecting Sets}

A variety of different set visualization techniques are available. Alsallakh et al.\ detail dozens of set visualization techniques in their review demonstrating how different techniques better match different needs of the user and data types. Certain techniques, such as Venn diagrams, are familiar and easy to interpret but are limited by the number of sets they are able to visualize (3 circles of a Venn diagram) \cite{bardou2014jvenn}. These simpler techniques do not work for our target user needs, as practitioners often have more than three models to compare. Other matrix based techniques, such as UpSet \cite{lex2014upset}, scale to many sets and allow for easy exploration of both sets and element attributes. Therefore, throughout the rest of this paper, we use UpSet style visualizations due to its scalability, clarity, and ability to handle complex data sets efficiently.

\begin{figure}[ht]
    \centering
    \includegraphics[width=0.9\textwidth]{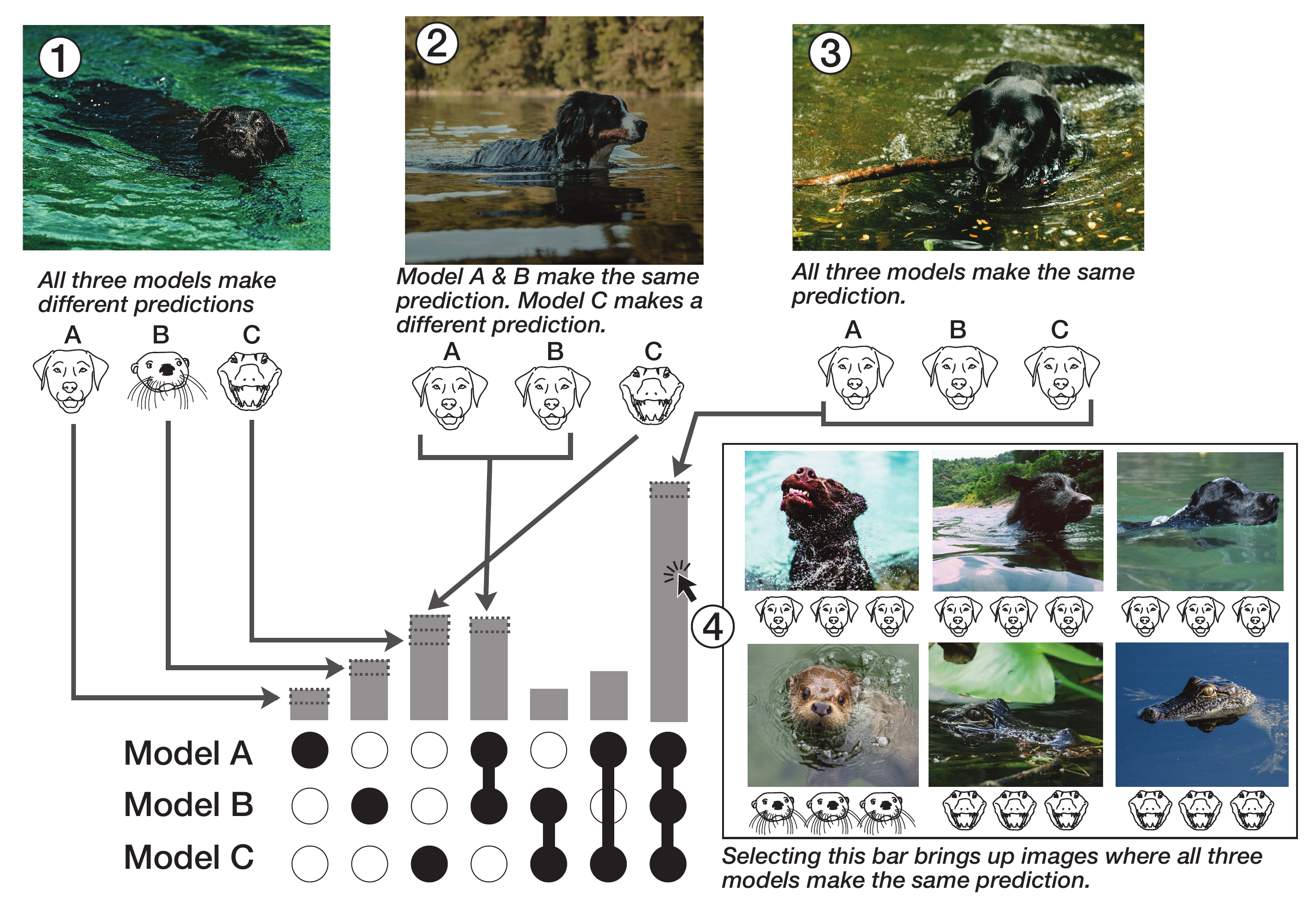}
    \caption{ Example of using UpSet visualization to compare model predictions. (1) All three models make different predictions for the same image. (2) Models A and B predict a dog, while Model C predicts a crocodile. (3) Set of predictions where all models agree, indicating a dog. (4) Selecting this set displays all images where all three models agree on the prediction of a dog, showing that models consistently identify a dog when not submerged in water.}
    \label{fig:foxUpset}
\end{figure}

To better demonstrate UpSet's utility in comparing models, consider a straightforward example involving image classification models. In Figure \cref{fig:foxUpset}, filled circles signify whether models agree or disagree on predictions. For example, if Model A and Model B are marked (filled circles) and Model C is not, it indicates agreement between Models A and B, with Model C having a different prediction (\cref{fig:foxUpset} (2) - Model A and B: Dog, Model C: Crocodile). Since Model C made a different prediction on this data instance, the bar where Model C is filled in and Model A and Model B are empty would also increase. Data instances may appear under several bars if models' predictions vary (\cref{fig:foxUpset} (1)). This overall view can give a glimpse of overall patterns. The proportionally taller bars over the Model A and B agreement set indicates that these two models generally make similar predictions. UpSet visualizations also facilitate exploration of why models agree on certain instances by analyzing element attributes. For instance, in the set where all models agree (\cref{fig:foxUpset} 4, all three circles filled in), it is evident that the models consistently agree on classifications when the image consists of the animals head.

This is just one simple example of how model outputs can be compared using set visualizations. In the following four case studies, we demonstrate how model outputs can be transformed into set type data for four real world ML applications. For each of these case studies, we include an accompanying notebook to generate the data, convert it to set type data, and generate the UpSet visualization in our supplemental materials (\textbf{TODO LINK}).

\subsection{Example 1: Clustering}
\label{case:clusters}
\begin{figure}[ht]
    \centering
    \includegraphics[width=0.95\textwidth]{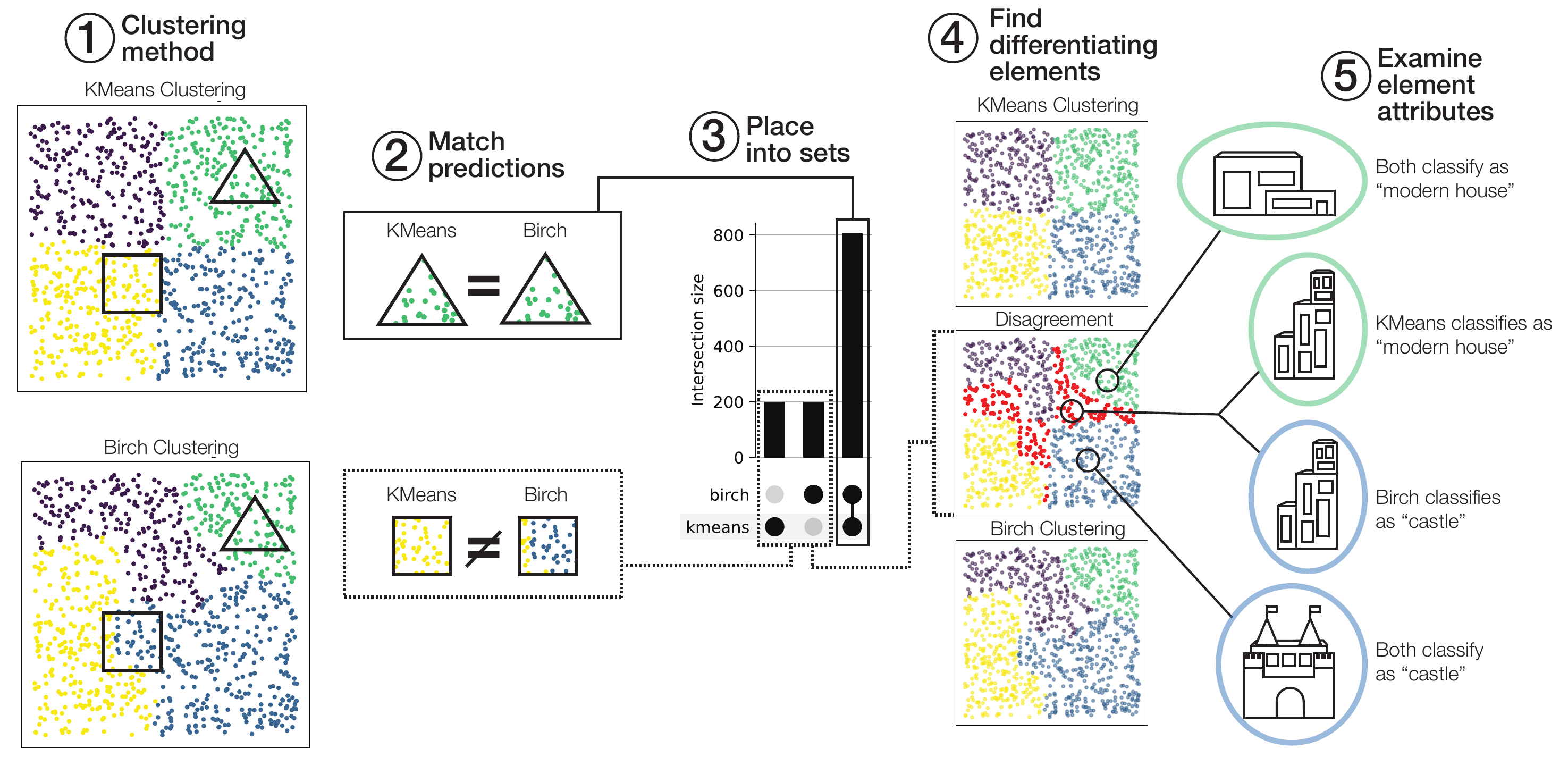}
    \caption{Transformation of data into sets based on cluster label agreement between K-Means and Birch algorithms. Clustering results are compared: (1) K-Means and Birch clustering results are matched. (2) Sets are generated based on agreement (triangles) or disagreement (squares) between the algorithms. (3) Differentiating elements are identified where K-Means and Birch disagree, highlighted in red. (4) Examination of element attributes reveals that houses with modern architecture align more closely with K-Means clusters. This set visualization helps isolate differences and understand model behavior.}
    \label{fig:clusterSet}
\end{figure}

One relatively simple example of how sets can be applied to explore data is by comparing different clustering algorithms. For example, a real estate analyst could be examining houses on the market by clustering them. After applying a dimensionality reduction algorithm to a set of house features, the analyst compares the outcomes of two different clustering algorithms, K-Means and Birch, using four clusters (\cref{fig:clusterSet} Cluster). The analyst can see overall that K-means and Birch make similar predictions so it is difficult to tell which one they should use.

To transform the data into set type data, they need a property of the data to match on. The analyst decides to match the different houses based on cluster label agreement (set generation rule). Simply, if the two cluster algorithms both place a house in the same cluster (same color in \cref{fig:clusterSet} (1)) then they go into the set of K-means and Birch agreement (\cref{fig:clusterSet} (2 - triangles)). If the algorithms label the houses differently (\cref{fig:clusterSet} (2 - squares)) then the prediction is added to the unique K-means and Birch sets. This is done for every point, creating sets of points that share the same label across algorithms. 

To quickly learn more about the two algorithms, the analyst finds the differentiating elements (\cref{fig:clusterSet} (4 - Find differentiating elements)). They are curious about which elements the models disagree on. The analyst isolates this subset of data and assesses whether the instances align more closely with the clusters formed by K-Means or those identified by Birch. Looking through the houses, the analyst finds that many houses the models clustered differently have a more modern architecture that more closely aligns with the K-means cluster choice (\cref{fig:clusterSet}  (5 - Examine element attributes - green cluster). Without sets, the analyst would have to manually search for all the points that differentiated the two models. Using sets allowed them to automatically isolate the interesting subset of the data and determine how the models were different.

\subsection{Example 2: Regression}
\label{case:regression}
\begin{figure}[!ht]
    \centering
    \includegraphics[width=0.98\textwidth]{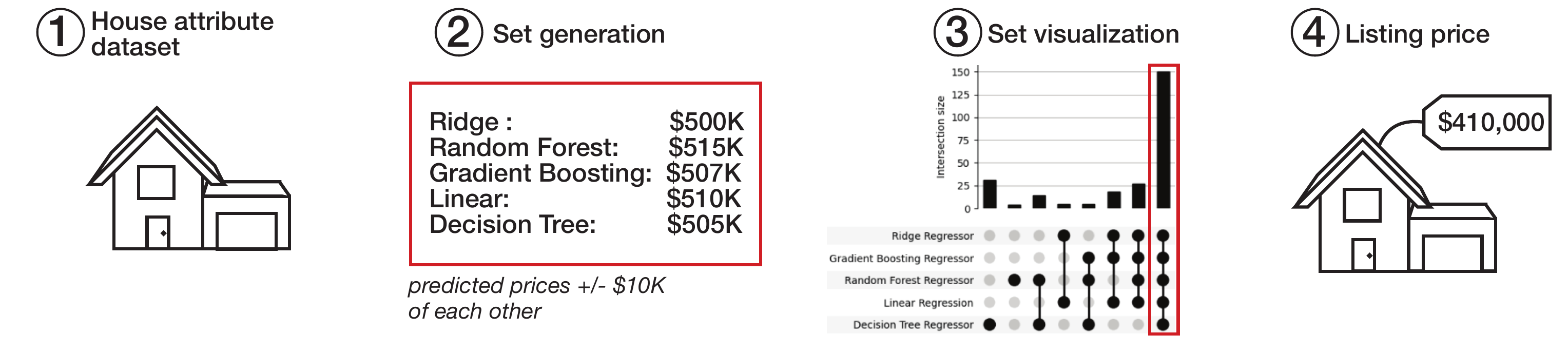}
    \caption{The figure shows how regression model predictions are converted into set-type data for house price estimation. In scenario A, predictions from five models (Ridge, Random Forest, Gradient Boosting, Linear, Decision Tree) for a house listed at \$410,000 are within \$10,000 of each other, indicating potential underpricing. 
    }
    \label{fig:houseSets}
\end{figure}

Let’s continue with the example of a housing dataset but instead look at how regression model predictions can be converted into set type data. The analyst has a dataset with attributes such as number of bathrooms, square footage, neighborhood and wants to predict the value of the house (\cref{fig:houseSets}). They want to determine which houses may be underpriced on the market but don't want to rely entirely on one model. They run five different regression models and obtained predictions from each model. The analyst again turns to sets to better compare the models. The analyst groups model predictions by considering those within a specified maximum distance (\$10,000)
as being in agreement, effectively applying a proximity-based rule to determine consensus (set generation rule) (\cref{fig:houseSets} (2)).

The analyst compares true listing prices with predicted data and identifies houses with significant price differences in the set where all five models agree (\cref{fig:houseSets} (3,4)). This model consensus suggests model accuracy (similar to ensemble methods \cite{dietterich2000ensemble}), indicating these houses are likely underpriced. Without using sets, the analyst would need to create lists of houses with the largest price differentials for each model, potentially highlighting model errors rather than accurate predictions.


\subsection{Example 3: Image Classification}
\label{case:imageClassification}

\begin{figure}[!h]
    \centering
    \includegraphics[width=0.9\textwidth]{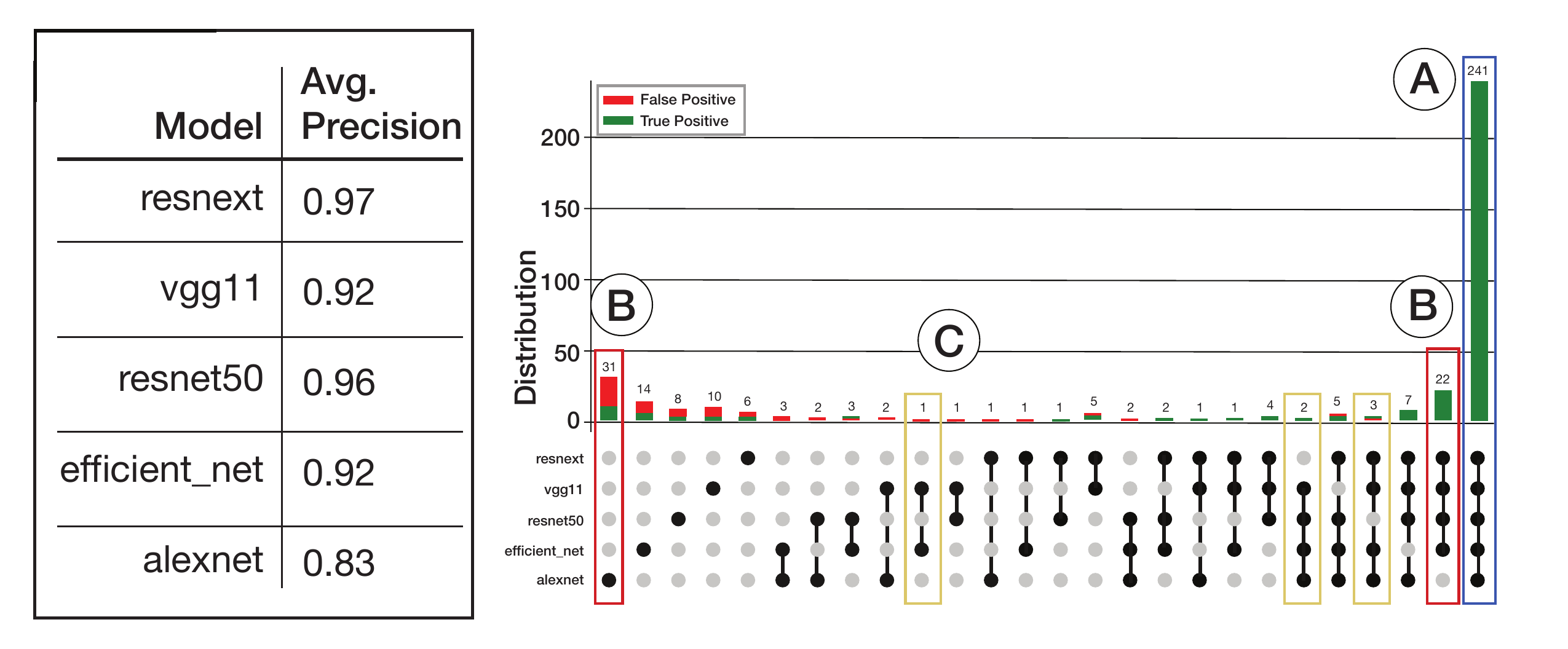}
    \caption{Comparison of multi-class image classification models using UpSet visualization. The left table lists five models with their average precision across all classes on a dog classification dataset. The UpSet plot on the right reveals various insights: (A) Models generally agree and perform well, as indicated by the tall right bar with all true positives; (B) AlexNet's unique predictions often result in false positives, unlike the other models; (C) Despite having the same average precision, VGG11 and EfficientNet rarely make the same predictions, as shown by the short bars and small agreement sets.}
    \label{fig:imageClassification}
\end{figure}
Set data and visualizations can also be used to compare multi-class image classification models. Similar to the example seen in \cref{fig:foxUpset}, an analyst wants to compare 5 different models on a dog classification dataset \cite{imagenette}.  They run 5 popular out of the box models (ResNeXt \cite{xie2017aggregated}, VGG11 \cite{simonyan2014very}, ResNet50 \cite{he2016deep}, EfficientNet \cite{tan2019efficientnet}, AlexNet \cite{krizhevsky2012imagenet}) on the test dataset. The analyst wants a better overall understanding of these different models' performance.

The model outputs are then converted into sets by matching on classification label agreement (set generation rule). An additional step is then taken to compare the sets against the ground truth, thereby classifying each set as a true or false positive. The analyst then gets an overview of the model performance by placing it into an UpSet visualization. Right away, they can see that these models generally agree and perform well (A - tall right bar with all true positives). Moving to a closer inspection, They see that Alexnet makes different predictions than the other models, and they are generally incorrect (B - False positives in unique Alexnet predictions). Finally, they see that while VGG11 and EfficientNet have the exact same averaged precision across all classes, they rarely make the same predictions (C - short bars, small agreement sets). Previously, the analyst could only see a ranked list of model performance without understanding if the models made similar or different predictions (\cref{fig:imageClassification} Left). The set visualization not only indicates model performance but also uncovers interesting patterns for further examination.

\subsection{Example 4: Object Detection}
\label{case:ObjDetection}

Our set visualization method and system development were driven by the needs of ML practitioners looking for better ways to parse and compare object detection model outputs. Object detection models identify objects in images by drawing bounding boxes around them and classifying the objects contained in the bounding box. Comparing their outputs is challenging due to the volume of data, with thousands of images and dozens of detections per image. Therefore, our collaborators encountered this problem and needed a better way to explore the model outputs.

\begin{figure}[h]
    \centering
    \includegraphics[width=.98\textwidth]{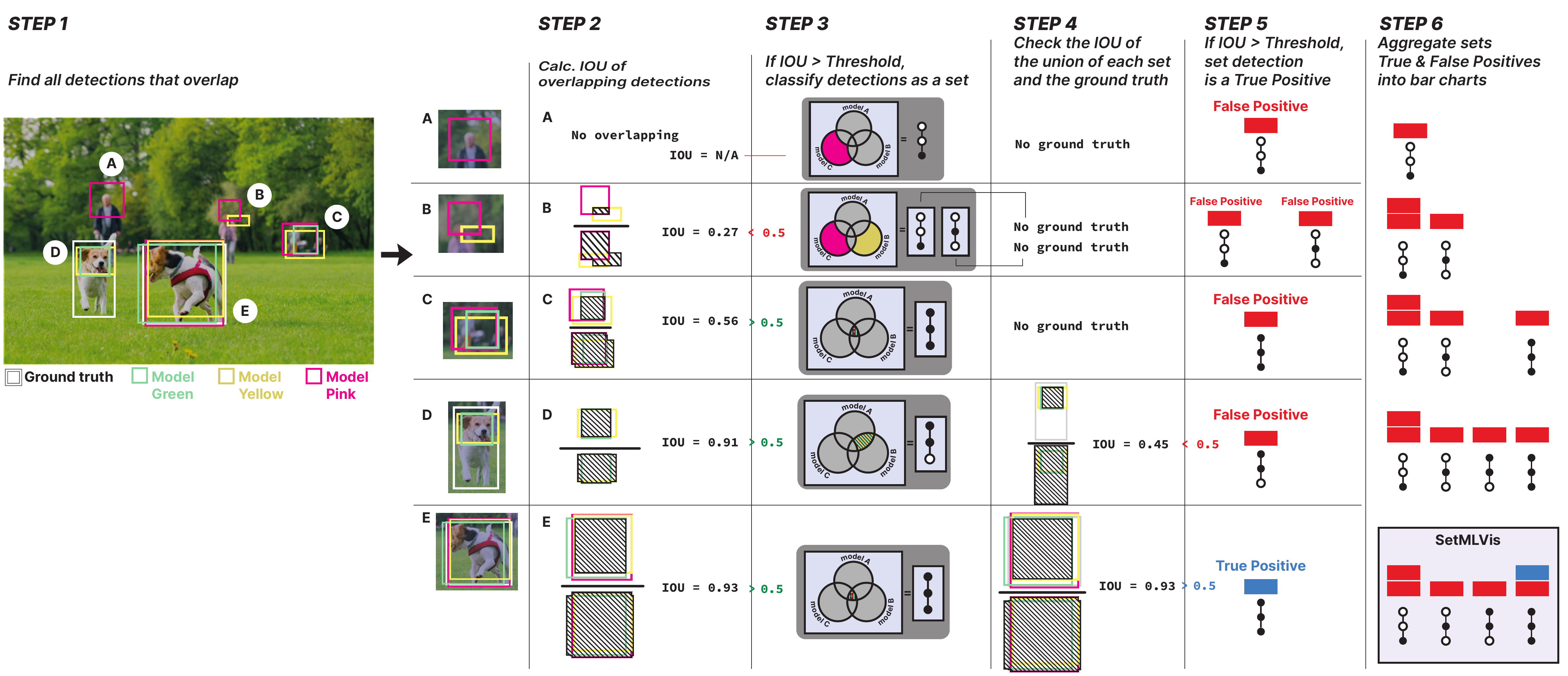}
    \caption{Overview of the method used in SetMLVis for comparing object detection models. The process involves identifying overlapping detections (Step 1), calculating Intersection over Union (IOU) for matched detections (Step 2), determining true or false positives based on IOU thresholds (Steps 3 and 4), and aggregating these results into visualizations for analysis (Steps 5 and 6).}
    \label{fig:setlogic}
\end{figure}

To demonstrate how our method can aid in comparing object detection models, we compare three different models' ability to locate and classify dogs in an image dataset. The data can be transformed into sets by aligning the bounding boxes from each model. A set is created if two or more bounding boxes have an intersection over union (IOU) of 0.3 or greater (Set generation criteria, \cref{fig:setlogic} step 2 and 3). This operates in much the same way as traditional evaluations where a model prediction is determined to be the same as the ground truth by getting the IOU between the two. In our case, we run this operation twice, first to get the sets (\cref{fig:setlogic} Steps 2 and 3) and then to determine if the prediction is a True or False Positive (\cref{fig:setlogic} - Step 4). 

Once the data is transformed and placed into set visualizations we can explore to (1) find which models are different, (2)  find the objects/images that cause models to make different predictions, and (3) identify the attributes of the objects the certain models struggle with. Looking at the image in \cref{fig:setlogic}, we can highlight several observations. First, we see that model pink makes the most unique predictions. These predictions are incorrect and (\cref{fig:setlogic} A, B) are often due to fuzzy background objects. The yellow and green models make similar predictions but make mistakes by localizing their prediction of a dog to just the face (\cref{fig:setlogic} D). All models correctly identify a dog when it is running at an angle (\cref{fig:setlogic} E). Using this information, we can better understand different element attributes affect model predictions. We use the same set generation logic and implement it in a user interface called SetMLVis to compare object detection models, as described in \cref{sec:setmlvis}.

\subsection{Sensemaking and Model Comparison}
These case studies demonstrate the variety of queries and type of information that can be gleaned using sets and set visualizations to analyze ML models. The analysis workflow using sets closely mirrors the general ML sense making framework established by Cabrera et al \cite{cabrera2023did}. In their framework, they describe the sensemaking process of ML models as (1) collecting instances and outputs, (2) generating schemas, (3) generating hypotheses, and (4) assessing those hypothesis \cite{cabrera2023did}. Using our method we (1) create multiple model predictions on the same inputs (collecting instances and outputs), (2) automatically generate sets by matching predictions (generating schemas), (3) generate hypotheses based on set relations and instance membership (generating hypotheses), (4) assess those hypotheses by looking at instance attributes (assessing hypothesis). Using sets neatly fits into this tested framework and allows practitioners to analyze their data in a workflow they are accustomed too.

\begin{table*}[!h]
\centering
\caption{Comparison of machine learning tasks (case studies) with and without set transformations}
\label{tab:ml_comparison}
\begin{tabular}{|p{1.7cm}|p{2.5cm}|p{3.5cm}|p{3cm}|p{3cm}|}
\hline
\textbf{ML Type} & \textbf{Question to be answered} & \textbf{Transformation to set data} & \textbf{Without Sets} & \textbf{With Sets} \\ \hline
Cluster  (\cref{case:clusters}) & \parbox[t]{2.5cm}{\raggedright Determine the best algorithm to find similar houses.} & \parbox[t]{3.5cm}{\raggedright Align clusters across models. Points with the same cluster label create sets.} & \parbox[t]{3cm}{\raggedright Parse through instance level data to find houses that different algorithms disagree on.} & \parbox[t]{3cm}{\raggedright Automatically extract subsets of houses that models disagree on.} \\ \hline
Regression (\cref{case:regression}) & \parbox[t]{2.5cm}{\raggedright Predict housing prices to find undervalued houses in the area.} & \parbox[t]{3.5cm}{\raggedright Predictions are placed into sets where each prediction is within a certain maximum distance from its neighbors.} & \parbox[t]{3cm}{\raggedright Look through the list of greatest differences in house price and actual price for all models. Cross-check with all other models to ensure validity.} & \parbox[t]{3cm}{\raggedright Use model agreements to quickly verify that the predicted price is realistic and not an outlier.} \\ \hline
Classification  (\cref{case:imageClassification}) & \parbox[t]{2.5cm}{\raggedright 	Compare models performance in identifying dog breeds.} & \parbox[t]{3.5cm}{\raggedright Each image is added to the set of models that agree on its classification.} & \parbox[t]{3cm}{\raggedright Use aggregate metrics such as Precision and Recall, which can hide underlying similarities and differences.} & \parbox[t]{3cm}{\raggedright Visualize specific prediction matches to see if models make the same predictions and if they are correct.} \\ \hline
Object detection  (\cref{case:ObjDetection}) & \parbox[t]{2.5cm}{\raggedright Identify instances that cause performance differentiation between models in detecting and localizing dogs.} & \parbox[t]{3.5cm}{\raggedright Calculate set IOU between model bounding boxes to determine if they are predicting the same object.} & \parbox[t]{3cm}{\raggedright Examine aggregate metrics and parse through thousands of detections, manually checking model prediction overlap.} & \parbox[t]{3cm}{\raggedright Automatically find differentiating subsets of bounding boxes and examine a smaller subset for clear insights.} \\ \hline
\end{tabular}
\end{table*}

\section{SetMLVis: An interactive tool for comparing object detection models}
\label{sec:setmlvis}

We go beyond simply proposing the use of set visualizations for model comparison by delivering SetMLVis, a fully functional tool that practitioners can immediately apply to complex model evaluation tasks. SetMLVis offers a practical, working solution to the challenge of comparing object detection models (\cref{case:ObjDetection})—a problem that has been inadequately addressed despite the proliferation of such models. Unlike existing tools that primarily focus on single-model exploration \cite{cabrera2023zeno, wang2022and}, SetMLVis enables direct comparisons between multiple models, leveraging set-based visualizations to provide clear insights into model behavior and performance. This system is fully integrated into standard workflows like Jupyter notebooks, making it easy for others to adopt and extend. SetMLVis fills a clear gap in the current landscape by offering a structured, interactive approach to a problem that has lacked adequate tools until now.

\subsection{How to Use}
From the start, SetMLVis was designed with usability and accessibility in mind. It integrates seamlessly with typical machine learning workflows, similar to established tools such as Zeno \cite{cabrera2023zeno}, AIFinnity \cite{cabrera2023did}, and Symphony \cite{bauerle2022symphony}. Available as an open-source tool on PyPI, SetMLVis is easy to install, integrate, and extend, making it widely accessible to users across different environments.

To use SetMLVis, users begin by selecting a folder containing the object detection model predictions they wish to compare. Next, they specify an Intersection over Union (IOU) threshold for set generation. The createSetJson function automatically processes the data by matching bounding boxes across models and converting them into set-type data (function process exemplified in \cref{fig:setlogic}, Listing \ref{lst:json_creation}). This data is then evaluated against the ground truth, and a JSON file is generated that contains both the sets and their evaluations, ready for use in the visualization tool.

The JSON file is then loaded into the SetMLVis visualization tool, which operates as a Jupyter widget, fully integrated within Jupyter notebooks. Once the data is in the visualization software, no further processing is needed, ensuring smooth data exploration. The entire workflow—from raw model output to interactive visualization—requires just two lines of code (Listing \ref{lst:json_creation}), allowing practitioners to easily deploy the software and stay within their familiar Jupyter environment.

\begin{lstlisting}[language=Python, caption={Example of generating JSON and initializing SetMLVis Widget in a Jupyter Notebook}, label={lst:json_creation}]
jsonfile = createSetJson(
    folderName="/path/to/your/folder/", # Folder with all model and ground truth predictions
    objectClass="yourObjectClass",     # Object class to be visualized
    jsonName="yourJsonName",           # Output JSON file for visualization tool
    setIou="predictionAggrementIOU"    # Set generation rule (IOU threshold)
)
w = SetMLVisWidget(data=jsonfile)      # SetMLVis widget
\end{lstlisting}

\subsection{Design}

From our review of studies on machine learning evaluation and discussions with collaborators specializing in object detection, we identified three key design goals for a model comparison system. These goals emerged consistently from both our conversations and the relevant literature. The design goals are as follows:

\begin{enumerate}
    \item \textbf{G1---Characterize overall model behavior:}  This is the general starting point in all ML comparisons. Individuals need to know the overall performance of the models to gain a general understanding, which helps in making initial assessments and guiding more detailed evaluations \cite{padilla2020survey}.
    \item \textbf{G2---Examine commonalities and discrepancies in model behaviour:} To compare items, we examine their similarities and differences. Cabrera et al.\ focus on sorting data at points where models show the most discrepancy, aiding users in generating hypotheses \cite{cabrera2023did}. By employing sets, data is automatically organized to not only reveal discrepancies but also highlight areas of agreement, thereby facilitating the creation of effective schemas for hypothesis generation.
    \item \textbf{G3---Investigate factors influencing model performance: } Understanding why and how models perform differently is crucial for confidently releasing a model. Using model failures to drive further design and improvements is a common iteration strategy. To achieve this, it is essential to analyze instance-level data attributes to understand the reasons behind model failures or successes \cite{suh2022visualization, raschka2018model, moore2023failurenotes} 
\end{enumerate}

As sets are often used for comparing different groups of objects, it is no surprise that set tasks closely align with these design goals. We find that the three types of set tasks outlined by Alsallakh et al.\ mirror our goals: tasks related to set and set relations (G1), tasks related to elements (G2), and tasks related to element attributes (G3) \cite{alsallakh2016state}. 

\subsection{Interface}

\begin{figure}[h]
    \centering
    \includegraphics[width=.8\textwidth]{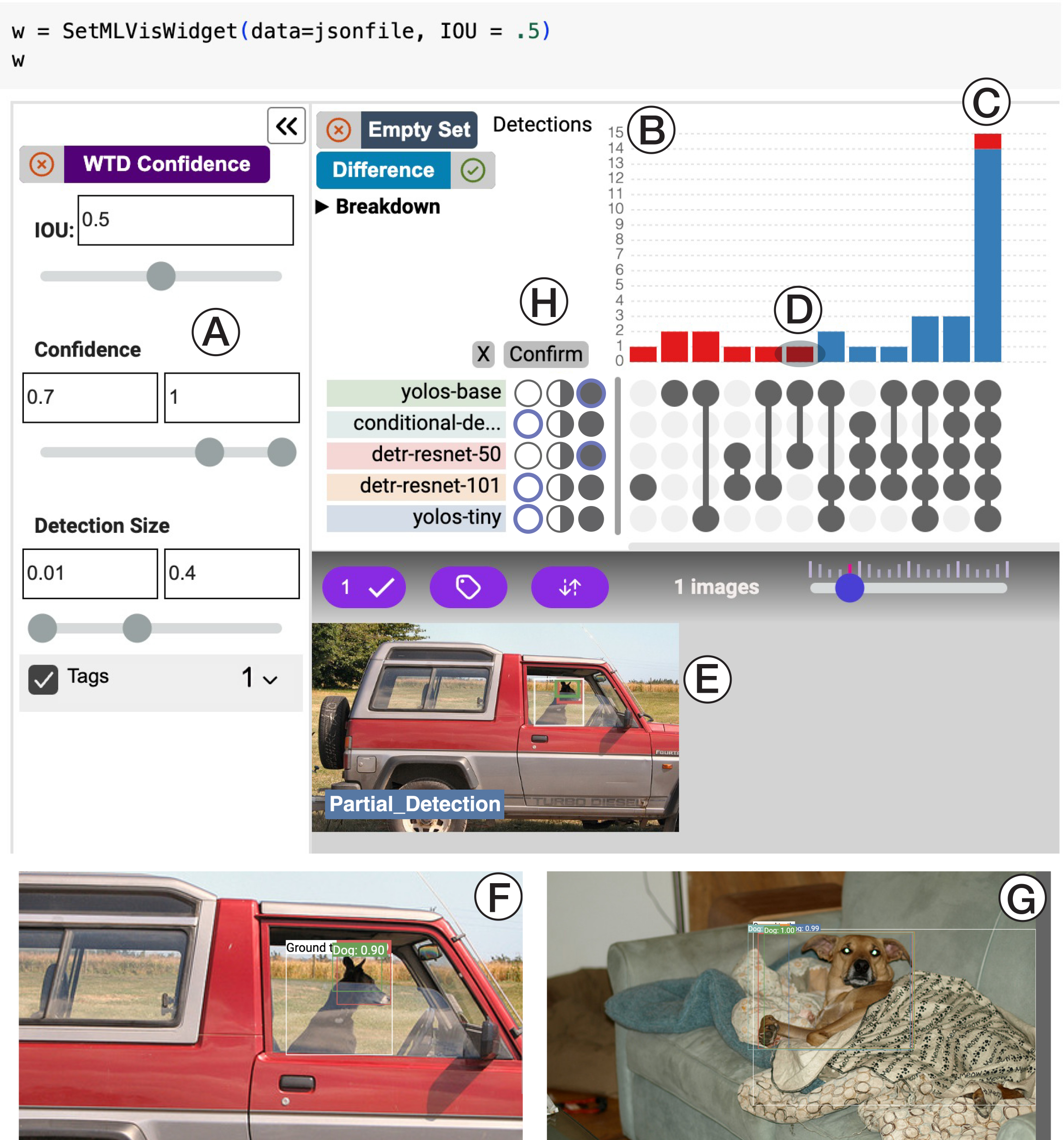}
    \caption{This image demonstrates the SetMLVis interface using the Imagenette dataset \cite{imagenette} and a range of popular open source models. An analyst can set the evaluation criteria to an IOU of 0.5 and detection confidence of 0.7 (A). The main visualization shows model prediction agreements and differences (B) using an UpSet style layout. They notice most models agree on predictions (C) but find that detr-resnet and yolo-base make similar errors (D). Clicking (D) reveals incorrect detections by these models (E). The detailed view (F) shows partial detections of the dog's head. Another set of predictions (G) (the false positives of bar (C)) shows a similar problem of partial detections highlighting occlusion issues. The analyst tags these images as ``Partial Detection'' and exports them for later analysis (E). (H) Demonstrates the interactive query system which can help filter the data to the desired subsets.}
    \label{fig:dogUpset}
\end{figure}

Using these design goals we designed and built our system, SetMLVis. The interface is designed to explore and compare object detection model predictions. While there are multiple components and views, the backbone of the data exploration revolves around a set visualization in the style of UpSet \cite{lex2014upset}. The set visualization allows for easy targeted exploration of model predictions. Adjusting criteria or selecting different sets allows the user to generate hypotheses about the models. This system offers a new way to examine model predictions, demonstrating the power of set-based analysis for uncovering patterns and differences in model performance.

\subsection{Usage Scenario}

Before breaking down the system components in depth, we will walk through a simple hypothetical use case to demonstrate the system's functionality. Let’s imagine our earlier scenario of an analyst wanting to compare several different models for detecting dogs. After running the models on the data and using the python function to parse it into set type data, they put the set data in the interface. 

Once the interface is open, they establish their evaluation criteria to an IOU threshold of 0.5 and a detection confidence threshold of 0.7 or above (A). Moving to the main visualization, they examine where different models make the same or different predictions (B). The data, parsed into clickable subsets, is displayed in an UpSet style visualization.

They notice that detr-resnet and yolo-base often agree but make incorrect predictions (C, Goal 2). Clicking on the bar at (C) brings up the specific detections where \textit{only} detr-resnet and yolo-base agree but are incorrect (E). Clicking on an image opens a detailed view (F), showing models detect only the dog's head, not the whole body (F, Goal 3), as seen in \cref{fig:dogUpset} (D). The same information can also be found using the query system (H). To find detections uniquely shared by detr-resnet and yolo-base, users would select the dark circles representing these two models (detection made), while selecting the white circles for other models (no detection). This functionality allows for targeted queries.

Next, they observe that while most models agree on predictions (tall bar for all model agreement, Goal 1), sometimes they are all incorrect. The user clicks this subset to identify challenging images (D, Goal 2). Examining the image attributes (Goal 3), they find the dog's head is occluded by a blanket, causing a partial detection. They tag both images (F, G) with a partial detection tag for later use and export (E).

All this analysis can quickly be done with a few simple clicks. Using SetMLVis, the analyst was able to (1) see that the models make similar predictions (tall right bar), (2) find that yolo-base and detr-resnet make similar errors (partial detections), (3) isolate dog images that are difficult for object detection models (partially occluded dogs), and (4) tag and export images for later development and use.

\subsection{System Components}

To create an effective analysis tool, we employ the multiple coordinated views (MCV) approach \cite{shneiderman2010designing}. Our system is composed of four interactive components, each designed to meet the identified design goals. Previous research of interviews with machine learning practitioners demonstrate a strong desire for more interaction between aggregate statistical outputs and instance-level data \cite{patel2008investigating, suh2022visualization}. We address this need by enabling users to click on bars in the Upset plot (aggregate view) to dynamically update the displayed images and bounding boxes (instance-level data) in the detail view. This intuitive interaction, combined with seamless integration into existing workflows, makes SetMLVis an effective tool for enhancing machine learning pipelines.

\textbf{Evaluation Criteria Widgets:} This panel (\cref{fig:dogUpset} (A)) includes simple sliders for users to adjust their evaluation criteria and filter out data quickly. 
The IOU slider determines the proportion of false positives to true positives, and the confidence slider filters out detections outside the given range of confidence scores. 
Changes to these widgets update the data shown in the Set Visualization panel.

\textbf{Set Visualization:} Shown in \cref{fig:dogUpset} (B), the set visualization is the focal point of the system as it displays the aggregate metrics and model similarities and differences (Goals 1 and 2). 
The primary visualization is a set visualization strategy similar to UpSet \cite{lex2014upset}. 
The process of generating the sets to be inputted into the system can be found in \cref{fig:setlogic}. 
The initial set visualization overview allows users to see the overall similarities and differences between the models. 
An interactive query system enables users to create meaningful data slices by isolating subsets of data to answer specific questions, such as ``where does model B detect something that model A does not?' (\cref{fig:dogUpset} (H))'. 
Users click the bars to display the relevant subset of images and model predictions in the Thumbnail View.
    
\textbf{Thumbnail view:} The thumbnail view (\cref{fig:dogUpset} (E)) allows users to preview the instance-level data (Goal 3). 
This area shows the images and the model bounding box predictions of the selected data slice. 
The user can quickly see which kind of images and detections caused issues for the different models \cite{cabrera2023did}. 
The user can define tags, apply them to different images, and access them in the next cell of the jupyter notebook. 
Tagging and exporting the images allows users to keep track of important errors and later extract them from the software to explain poor model performance and iterate on the model through failure-driven design \cite{cabrera2023did, moore2023failurenotes}. 
If the user needs to see the detection in more detail, they can open up the detail view of the image by clicking on any thumbnail.
    
\textbf{Detail View:} The detail view shows the instance-level data and allows users to see all the detections the models make on a specific image. The larger view allows users to closely examine the element attributes (Goal 3, \cref{fig:dogUpset} (E, F)). 
Examining each image can help users generate and test hypotheses about why certain models perform the way they do.

\section{Evaluation}

For our initial system evaluation, we conducted a mixed-methods comparative study between SetMLVis and FiftyOne \cite{moore2020fiftyone}, a widely respected open-source software for object detection model evaluation. Our goals were twofold: 1) to compare SetMLVis with state-of-the-art object detection visual analytics systems, and 2) to examine how set visualization impacts user performance on model comparison tasks. While FiftyOne offers many similar features—such as a comparable interface, tools for adjusting evaluation criteria, visualizing bounding box overlays, and user-friendly image views—it lacks the set visualization feature central to SetMLVis (\cref{tab:ui_features}). We do not aim to discredit FiftyOne, which excels in many areas, particularly in model exploration, that fall outside the scope of our analysis. Instead, by focusing only on the comparable user interface elements, we aim to reduce extraneous variables and better isolate the impact of set visualizations on model comparison performance.

\begin{table}[h!]
\centering
\begin{tabular}{|l|c|c|}
\hline
\textbf{User Interface Feature}                   & \textbf{SetMLVis} & \textbf{FiftyOne} \\ \hline
Sliders to adjust evaluation criteria             & \checkmark        & \checkmark        \\ \hline
Set visualization for detection exploration       & \checkmark        &                   \\ \hline
Thumbnail view of images with bounding boxes      & \checkmark        & \checkmark        \\ \hline
Detail view for element attribute inspection      & \checkmark        & \checkmark        \\ \hline
\end{tabular}
\caption{Comparison of User Interface Features between SetMLVis and FiftyOne}
\label{tab:ui_features}
\end{table}

In our quantitative analysis, we do not explicitly attribute any observed differences in performance directly to the use of set visualization. While causality cannot be definitively established, any observed differences may suggest the potential impact of set visualization. To help provide more depth and nuance to the quantitative results and gain clarity on why the performance was different, we also ask participants to think aloud during portions of our study. We hope that a combination of these methods will indicate if sets have an influence on model comparison tasks and provide motivation for future work in this area. To that end we hope to answer two research questions:

\begin{enumerate}
    \item Compared to FiftyOne (traditional visualization techniques), does SetMLVis (set visualizations) enhance participants' ability to correctly compare object detection models across a range of tasks (\cref{tab:system_goals_evaluation})?
    \item  Does SetMLVis (set visualizations) reduce cognitive workload in model comparison tasks compared to FiftyOne (traditional visualization techniques), as measured by NASA TLX \cite{hart2006nasa}?
\end{enumerate}

\begin{figure}[h]
    \centering
    \includegraphics[width=.95\textwidth]{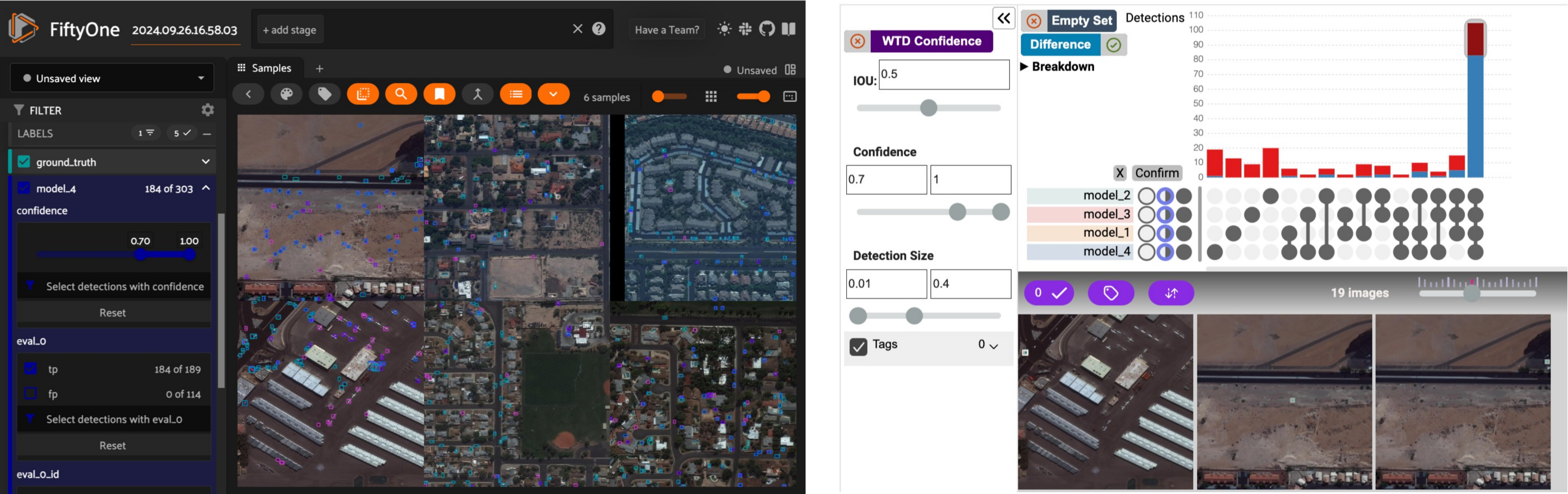}
    \caption{The two interfaces compared in the study.
    FiftyOne is on the left and SetMLVis is on the right.}
    \label{fig:interfaceComparison}
\end{figure}

\subsection{Participants}

The comparative study was conducted with 8 graduate students (5: M, 3: F) majoring in computer science at a large university in the northeast United States. There was no specified eligibility criteria for participation as the tasks did not demand specialized knowledge in object detection beyond the introductory information provided at the beginning of the study. The study took around 2 hours and participants were compensated with \$50 gift cards for their time. We chose to run our within subjects experiment with 8 participants because the 2 hour experiment time made recruitment difficult and required substantial experimental effort.  We chose our sample size because it aligns with previous studies \cite{moore2023failurenotes, cabrera2023did}, and our initial (though incorrect) power analysis suggested that only five participants were needed. A subsequent, corrected power analysis indicated that more participants were required, but despite this, we still observed substantial differences between the two systems in terms of task accuracy and cognitive load.

\subsection{Methodology}
For our study, we asked participants to complete five tasks on both systems. While these tasks do not encompass all possible machine learning evaluation tasks, they represent a diverse subset of tasks that practitioners may encounter in a model comparison workflow. Each task is grounded in the relevant literature and linked to common model comparison goals (detailed information on each task can be found in Table \ref{tab:system_goals_evaluation}). Before the study began, we provided background information on object detection model evaluation, video tutorials on the essential UI features of both systems, and 5 minutes of exploration on both systems (study procedure can be found in the supplemental materials: \url{https://osf.io/afksu/?view_only=bb7f259426ad425f81d0518a38c597be}).

During the study, participants were given 5 minutes to complete each task. This time frame was established based on pilot studies, which indicated that it was a reasonable duration for task completion with at least one of the systems. While it is possible to find answers for all tasks with unlimited time, limiting the time to five minutes helped ensure the study remained practical and reflective of real-world scenarios where users do not have an entire day to find answers.

The counterbalancing strategy employs a Latin square design, where task order and system exposure are rotated based on participant number to minimize order effects. For each participant, the starting task is shifted in a cyclic manner, and system assignment is alternated within tasks. This method ensures that no participant consistently experiences the same task-system pairing, thereby mitigating potential confounds such as learning effects, task fatigue, or system familiarity. Since there are many possible answers for each task, the correctness of each response was determined by the study administrators (primary and third author). If the participants could not answer the question before the 5 minutes were over or only provided incorrect answers during the allotted time, the task was marked incorrect for that system. 

Finally, after \textit{each} task for \textit{each} system, a 7 point NASA-TLX survey was administered \cite{hart2006nasa}. Collecting NASA-TLX data for each task and each system allowed for an examination of not just overall system performance but also what specific types of tasks might most benefit from set visualizations. While participants answered the questions, we encouraged but did not require them to think aloud and explain their responses. In our results illustrative quotes are used to demonstrate both the challenges and benefits of both systems.

\begin{table*}[tbp]

\caption{The table shows the tasks performed by the participants of the study and how they connect to system design goals and are grounded in real model evaluation practices.}
\label{tab:system_goals_evaluation}

\begin{tabular}{p{0.05\linewidth}p{0.2\linewidth}p{0.35\linewidth}p{0.35\linewidth}}
    \hline
    \textbf{Task} &\textbf{System goal} & \textbf{User task} & \textbf{Model comparison goals found in literature} \\
    \hline
   1& G1: Characterize overall model behavior & Which two models share the most detections, both correct and incorrect? Show evidence to support this. & "However, in order to take advantage of models with complementary strengths, we also explicitly encourage diversity by pruning away models that are too similar to previously selected models." \cite{huang2017speed} \\
    \hline
    2&G2: Examine commonalities and discrepancies in model behavior & Find an instance where two models correctly detect a car (True Positive) while the other two fail to detect it. & Even at the beginning stages, it’s still beneficial for our team to know what the model is all about. What are the strengths of this model? \cite{suh2023metrics} \\
    \hline
    3&G2: Examine commonalities and discrepancies in model behavior & Pick one model and find instances on two different images where that model makes an incorrect prediction that no other model does. & "SMEs need comprehensive details for a model’s weaknesses at all stages of development; data scientists need context for why a model is underperforming." \cite{suh2023metrics} \\
    \hline
    4&G3: Investigate factors influencing model performance & Pick a model and provide a unique reason that it is generating false positives. Provide three false positive detections and describe a pattern or connection between them. & "A hypothesis is a high-level description of a behavior (e.g., the AI fails in low light, or the AI works best for long sentences) along with supporting evidence." \cite{cabrera2023did} \\
    \hline
    5&G3: Investigate factors influencing model performance & Find three objects and describe their attributes that cause all four models to make an incorrect prediction. & "Point to outliers in the model’s performance and data space with known or plausible explanations." \cite{suh2023metrics} \\
    \hline
\end{tabular}
\end{table*}

\subsection{Analysis}
Our study and analysis were pre-registered and can be found in our supplemental materials. To address RQ1, we used a McNemar test \cite{mcnemar1947note} to statistically compare per-participant differences in task accuracy across both systems (FiftyOne and SetMLVis). We first converted answers into binary values (correct/incorrect) and created a contingency table for each participant and task. We then aggregated results across all tasks and participants to assess overall differences in accuracy between the two systems. 

For RQ2, we analyzed cognitive workload using a Wilcoxon signed-rank test \cite{wilcoxon1992individual} on the NASA-TLX scores \cite{hart2006nasa}. Each participant rated cognitive workload for each task on both systems across several dimensions (e.g., mental demand, effort). We aggregated these ratings by averaging them across all tasks for each system, producing a single NASA-TLX score per participant per system. The Wilcoxon signed-rank test then compared these paired scores to assess cognitive workload differences between FiftyOne and SetMLVis. Additionally, we conducted visual analysis to gain more granular insights.

Illustrative quotes from the think-aloud protocol are used to provide context and highlight key points of participants' experiences. These quotes highlight both positive and negative aspects as experienced by participants, providing deeper insights into the tasks for which set visualizations are particularly suited or unsuited. Although we didn’t conduct a comprehensive qualitative analysis, we believe that using illustrative quotes can still provide valuable context. These quotes highlight specific strengths and weaknesses of each system that may not be fully captured in the quantitative data.

\section{Results}

We found strong evidence that participants performed these tasks more accurately using SetMLVis than FiftyOne. 
The McNemar test comparing participants' overall task accuracy between SetMLVis and FiftyOne indicates that a difference is likely ($p = 0.004, \chi^2(1, N = 35) = 0.0$).
\cref{tbl:correct_answers} shows the overall percentage of correct answers for each visualization and task.
In Tasks 1--4 SetMLVis has a higher percentage of correct answers while all participants answer Task 5 correctly with both systems.

SetMLVis consistently outperformed FiftyOne across all tasks, with a particularly large advantage in Task 1, where participants using SetMLVis answered correctly 37.5\% of the time, compared to 0\% when using FiftyOne. Task 4 also showed a substantial difference, with SetMLVis achieving 100\% accuracy, while FiftyOne reached only 50\%. The task-specific accuracy results, combined with the general findings of the McNemar test, indicate that SetMLVis generally improves participants' accuracy on model comparison tasks. These findings suggest that the SetMLVis visualization approach may lead to a better understanding of the models' strengths and weaknesses.

\begin{table}[ht]
\centering
\renewcommand{\arraystretch}{1.2} 
\begin{tabular}{|p{0.45\linewidth}|r|r|}
    \hline
    \textbf{Task Abbreviation (Task Number)} & \textbf{FiftyOne (\%)} & \textbf{SetMLVis (\%)} \\ \hline
    Model Similarity (1)                     & 0.0   & \textbf{37.5}  \\ \hline
    Car True Positive Comparison (2)         & 87.5  & \textbf{100.0} \\ \hline
    Unique Model Errors (3)                  & 87.5  & \textbf{100.0} \\ \hline
    False Positive Pattern Identification (4)& 50.0  & \textbf{100.0} \\ \hline
    False Positive All Models (5)            & 100.0 & 100.0 \\ \hline
\end{tabular}
\caption{Percentage of correct answers per task for FiftyOne and SetMLVis (higher is better). Participants were more accurate using SetMLVis for Tasks 1--4 and were always correct with both tools for Task 5.}
\label{tbl:correct_answers}
\end{table}

\begin{figure}
    \centering
    \includegraphics[width=0.5\linewidth]{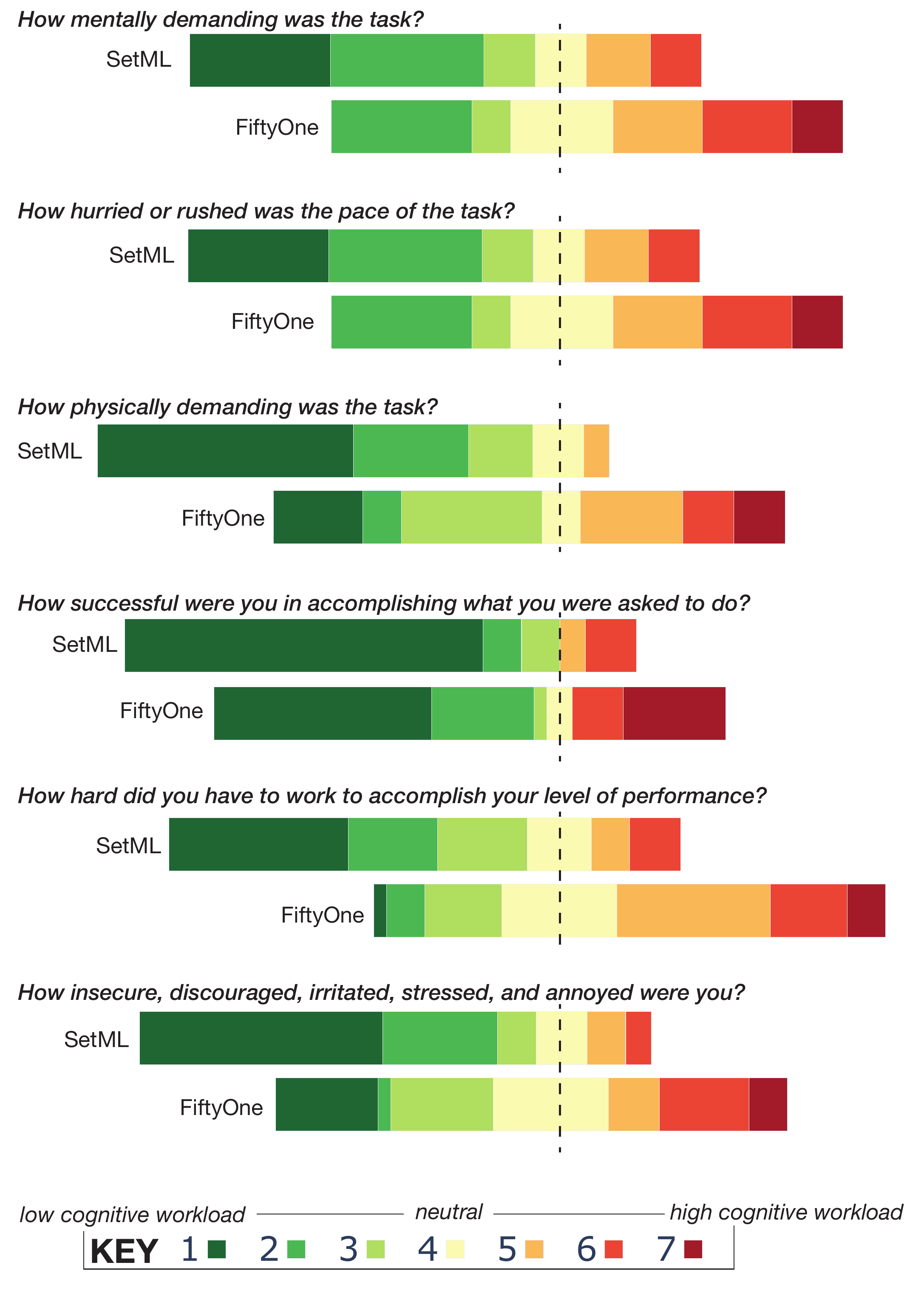}
    \caption{Comparison of NASA-TLX cognitive workload ratings between SetMLVis and FiftyOne across several dimensions: mental demand, physical demand, effort, and overall workload. Lower scores represent lower cognitive workload. Participants consistently reported lower cognitive workload when using SetMLVis, particularly for mental and physical demand.}
    \label{fig:likert-results}
\end{figure}

The Wilcoxon signed-rank test showed a likely difference in cognitive workload between the systems ($W = 1279.0$, $p < 0.001$). Across all NASA-TLX dimensions, SetMLVis resulted in lower reported workload compared to FiftyOne (\cref{fig:likert-results}). The most substantial improvements were seen in mental and physical demand, as well as effort, with mental demand showing the greatest difference, based on the largest mean score difference between systems for this dimension. This suggests that SetMLVis' visualization techniques reduced the cognitive load required to complete the tasks and enabled participants to more efficiently retrieve answers through the interface. Task 2 (Find an instance where two models correctly detect a car (True Positive) while the other two fail to detect it) had the greatest difference in cognitive workload, determined by the largest cumulative NASA-TLX workload difference across the systems.

Overall, the results indicate that SetMLVis not only improved task accuracy but also reduced cognitive workload suggesting that set visualizations allow for a more efficient and user-friendly method for comparing object detection models.

\section{Discussion}

\subsection{SetMLVis Improves Accuracy---Especially for Complex Tasks}
Task accuracy improved across all tasks except Task 5 (False positive all models), where participants performed equally well with both models. The greatest differences were found in tasks with greater complexity, which included those requiring more involved visual inspections or those where reaching the correct answer demanded more reasoning steps.

The differences in task accuracy between the two systems were particularly noteworthy for Tasks 1 and 4. For Task 1 (Model similarity), the challenge largely stemmed from the manual effort required in FiftyOne. To find the correct answer, participants had to go through each image and count all shared detections between each combination of models. As participants worked through the task, they often realized that the aggregate metrics for individual models were not useful in determining the similarity between them: \textit{"Even though they have the same true positive number, how do I know that they are the same true positives?"} (P5). With hundreds of detections and no way to make the process more efficient, many participants realized that completing this task in the allotted time was near impossible. One participant remarked, \textit{"In FiftyOne, it’s much more manual. You can’t query the system, so you just have to go through each image. That takes forever."} (P6). 

While more participants (3/8) correctly completed this task using SetMLVis, the overall performance was still low. The difficulty arose from understanding what the neutral buttons in the query system (\cref{fig:dogUpset} (H)) meant. Participants often looked for total matching detections by querying agreement between two models while also querying where the other two models did not make detections, leaving them with a subset of data that found unique shared predictions instead of total shared predictions. While, there was confusion with the neutral buttons, some participants began to understand how they worked after exploring the system for longer. One participant noted, \textit{"I had trouble figuring out how the neutral buttons worked. At first, I didn’t know what they were doing until I started playing around with the detail view and saw their impact."} (P4).

Task 4 (False positive pattern identification) was difficult because it required more complex reasoning to solve. The participants had to not only locate objects but also identify patterns among them. This process mimics how developers hypothesize about model behaviors, assess, and report on them \cite{cabrera2023did}. While participants could find unique incorrect detections in FiftyOne (Task 3), they struggled to build off this task as they were required to find where the incorrect detections occurred, which model made them, and whether multiple similar errors existed. SetMLVis eliminated much of this mental demand by presenting the correctly selected subset to help participants easily find patterns. One participant commented, \textit{"With SetMLVis, I didn’t have to constantly switch views or manually filter. The query system already filtered out what I didn’t need, so I could focus on just the patterns that mattered"} (P5).

\subsection{SetMLVis Excels at Model Comparison but is Missing Features for Single Model Exploration}

Participants consistently stated that SetMLVis was more useful when comparing multiple models. Matching detections and placing them into clickable sets made it easier to isolate commonalities and discrepancies, facilitating model comparison. In contrast, FiftyOne required participants to frequently toggle between hiding and showing different detections from models, which led to both physical and mental fatigue. One participant noted, \textit{"There was a lot of clicking in FiftyOne; it was physically demanding, especially having to manually adjust the settings to see the results."} (P2). This issue is reflected in the NASA-TLX scores, where mental demand, physical effort, and overall effort were substantially higher for FiftyOne (\cref{fig:likert-results}). This fatigue resulted from participants manually tracking and clicking to replicate what the set generation algorithm does in SetMLVis. As one participant noted, \textit{"Mentally, FiftyOne required more effort because you’re constantly tracking and comparing detections. SetMLVis did all that for me."} (P7).

In particular, Task 2 (Car true positive comparison) had the highest difference in cognitive workload between the two systems. It required a lot of guessing and checking in FiftyOne to see if certain objects were shared across models. One participant explained, \textit{"In FiftyOne, I had to manually go to an image and visually inspect where models matched. It added an extra layer of verification."} (P5). Additionally, some combinations of models yielded no correct answers, which participants could only verify by manually inspecting all images. As one participant pointed out, \textit{"Using FiftyOne felt like a guessing game because it wasn’t clear which objects were shared. I had to go through every image."} (P6). SetMLVis made this task easier and quicker, as users could instantly see whether a valid data subset existed in the set visualization.

However, despite SetMLVis’ strength in model comparison tasks, many participants found FiftyOne to be superior for single model exploration. FiftyOne’s focus on showing all detections in the main view made it easier for participants to get an overall picture of a single model’s performance. The option to toggle between true positives and false positives in FiftyOne, which was not available in SetMLVis' detail view, was appreciated by users. \textit{"With FiftyOne, it was easier to see a single model’s results on the whole image, especially when toggling false positives and true positives."} (P3). Another participant added, \textit{"The FiftyOne one, I think, is better for if you want to compare multiple detections from the same model... SetMLVis is probably better if you want to compare different models, like figuring out where one model failed while the others didn’t."} (P7).

\subsection{Query System Allows for Intuitive Question Answering}

A surprising finding is that nearly every participant used the query system (see \cref{fig:dogUpset} H) as the primary way of interacting with SetMLVis to find their answers. Once users understood how the query system worked, the tasks became straightforward and quick to complete. As one participant mentioned, \textit{“Once I had the confidence that I know how to use the query feature, and it does work as expected... the final task was like a split second job.”} (P6)

The success of the query system may stem from its design, which allows users to efficiently retrieve the data needed to answer the tasks. The main upset style visualization was used less, with participants primarily using it when participants they wanted to see where all models agreed (see \cref{fig:dogUpset} D) or where only one model made a prediction. Even in these cases, users sometimes relied on the query system, isolating a single bar which could have been clicked directly. \textit{“I used the query system to see what models 1 and 2 got right and wrong. I didn’t really need the big chart; just clicked the filters to show me what I wanted.”} (P7) Rather than scanning all possible set combinations in the full UpSet-style visualization, participants accessed the relevant data directly by aligning their reasoning steps with button clicks. \textit{“If I clicked the dark circle for models 1, 3, and hit confirm... it shows me true positives and false positives shared between those models.”} (P3)

While the query system was effective, it benefited from being paired with the UpSet visualization. Seeing the full UpSet visualization helped users better understand how the query system worked and provided immediate feedback on how adjustments, like IOU or confidence, influenced the data. This combination allowed users to visually confirm their query results and understand the broader context of the data. Another reason participants may have preferred the query system over the UpSet visualization is that the tasks were more focused on precise querying rather than broad pattern recognition, which limited the need for scanning across multiple set intersections. Nevertheless, systems looking to leverage the power of set visualizations without requiring complex interfaces should strongly consider adding a query system to enhance functionality while saving space.

\section{Limitations/Future work}

\subsection{Limitations}

Although our study provides meaningful insights into system performance and visualization strategies, several limitations must be acknowledged. These factors affect the generalizability of our findings and highlight areas for improvement.

First, the participant pool was small, limited to eight graduate students. While we observed substantial differences between the systems, a larger and more diverse sample would improve the study's power and generalizability. The limited population prevents us from stating that set visualizations are universally beneficial but our participants likely represent future users of these systems.

Participants also faced a learning curve, particularly in the tasks they faced first. Some complex tasks introduced at the beginning may have been overwhelming for some participants and affected our results despite our counterbalanced design intended to minimize this effect. Additionally, the study's scope was narrower than real-world scenarios. We limited participants to six images with around 40 detections per image, while practitioners typically work with much larger datasets. This simplification ensured manageability but may have limited the realism of the task environment.

Users also broadly commented on three specific aspects of SetMLVis that posed challenges. First, zooming into smaller detections was difficult, particularly in the notebook environment where the detail view lacked a clear zoom feature. Second, while participants enjoyed being able to toggle true and false positives in FiftyOne’s detail view, they could not find a similar feature in SetMLVis. Lastly, SetMLVis does not currently support the display of false negatives in the set visualization. Participants with experience in object detection evaluation emphasized that false negatives are an important part of model evaluation and expressed interest in seeing this functionality added in the future.

Beyond the system-specific issues, there are broader challenges related to set visualizations themselves. Proper documentation and onboarding are essential, particularly for understanding the query system. During the exploration phase with SetMLVis, participants had questions about what data was being displayed and were sometimes mistaken during the study, as seen with the low accuracy scores for Task 1. If users don’t fully understand how the set generation algorithm works or what is being visualized, they can reach incorrect conclusions, even though the system is designed to simplify the comparison process.

Another challenge is deciding on set generation criteria, which can lead to varied results. For instance, if the set IOU is too high, it may seem as though models are making different detections when they are not. On the other hand, if the set IOU is too low, models may appear to agree more than they should. It's important to keep in mind that the set IOU should generally be lower than the ground truth IOU since its purpose is to match predictions rather than evaluate their accuracy. While users familiar with evaluating machine learning models may be able to adapt to creating effective set generation criteria, it still involves a learning curve to align the criteria with user intuition.

\subsection{Future Work}
Our system demonstrates how set visualizations can be effectively used to explore object detection models, but the potential applications extend far beyond this domain. In our case studies, we briefly highlight how set visualizations could assist in a broader range of machine learning tasks. Future work should aim to build on this by developing visual analytics systems that facilitate model exploration across various machine learning contexts, expanding the utility of set visualizations.

Future work should also explore how set visualizations can enhance understanding of hyperparameter tuning and model iteration. By visualizing the consistency of predictions across different hyperparameter configurations, practitioners can gain clearer insights into where tuning reduces errors or introduces variability. This offers a more intuitive way to track how specific adjustments are impacting model behavior and helps determine whether further refinements are necessary.

Set visualizations also allow practitioners to identify persistent issues, such as recurring false positives and false negatives, across multiple iterations. This helps pinpoint which hyperparameters are not effectively improving model performance, guiding more targeted tuning efforts. Incorporating set visualizations into the model development process could streamline tuning by making the effects of hyperparameter changes more transparent and actionable.

The use of set theory also opens up new possibilities for evaluating models, going beyond traditional aggregate metrics. While these metrics provide a starting point, they cannot capture whether models are making the same predictions. By applying set mathematics, we can develop metrics that more effectively assess prediction similarities between models. We introduce two metrics here as a starting point. First, the Jaccard similarity metric \cite{jaccard1901etude}, which provides a quick overview of prediction overlap between models, and second, the Tversky similarity index \cite{tversky1977features}, which expresses the extent to which one model’s predictions are contained within another. While these examples show promising results, we believe there are more metrics to be explored.

\subsubsection{Jaccard Similarity Metric}

\begin{figure}[h]
    \centering
    \includegraphics[width=.9\textwidth]{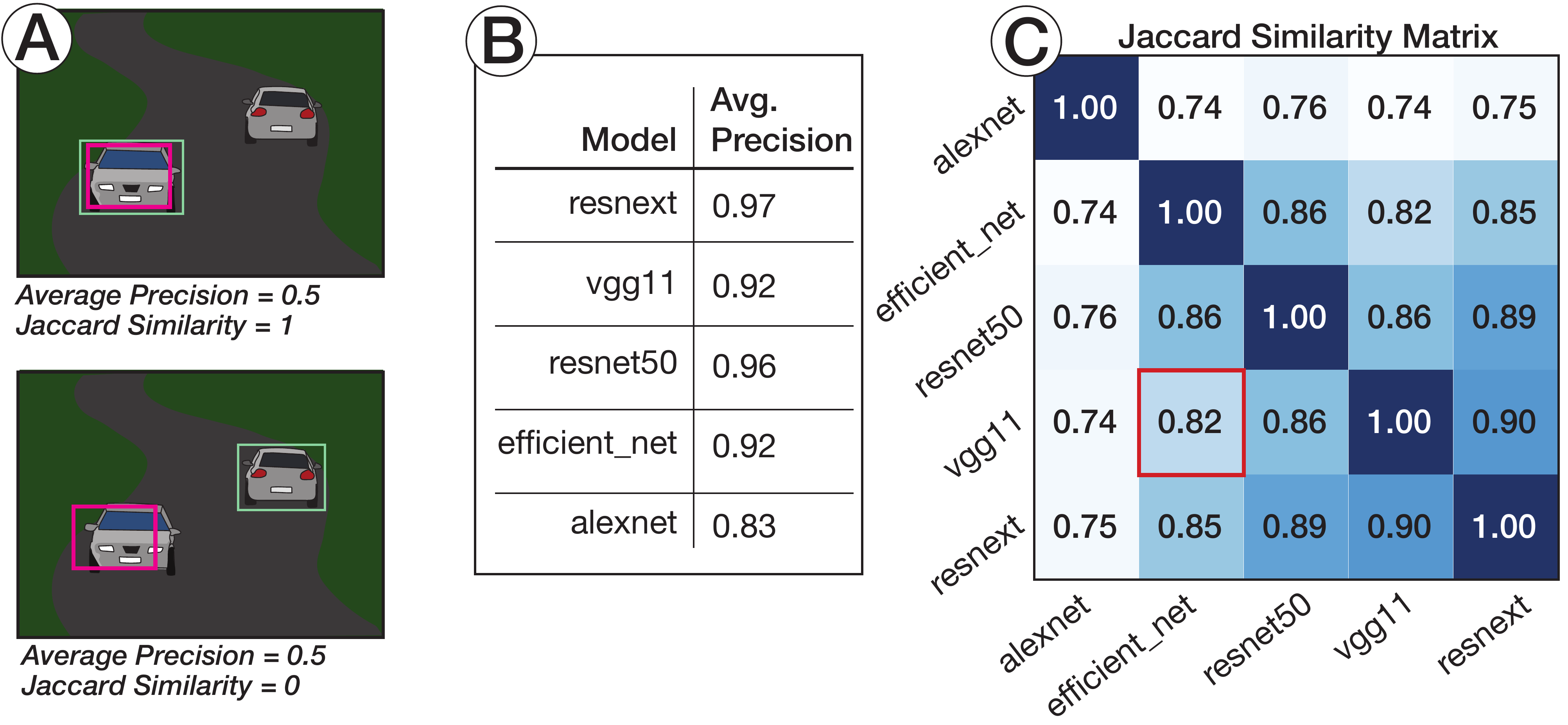}
x    \caption{Panel A shows two models with the same average precision (AP = 0.5) but different prediction overlaps, with Jaccard similarity values of 1 (top) and 0 (bottom). Panels B and C illustrate the Jaccard similarity matrix and average precision for five models, revealing that, despite similar average precision, VGG11 and EfficientNet have a low Jaccard similarity score (0.82), indicating less overlap in their predictions.}
    \label{fig:jaccard}
\end{figure}

\Cref{fig:jaccard} A demonstrates how the Jaccard similarity metric can complement metrics such as average precision. The Jaccard similarity metric measures similarity between two sets by dividing common elements by the total elements. The resulting value ranges from 0 (no similarity) to 1 (perfect similarity). When two models, Pink and Green, have the same average precision, the Jaccard metric can quickly discern if those models make the same predictions (\cref{fig:jaccard}A - top) or different predictions (\cref{fig:jaccard}A - bottom). This metric is easily interpretable and can be scaled to more models and datasets and shown visually through a heatmap (\cref{fig:jaccard} B). In \cref{fig:jaccard} B, we can see similar conclusions to \cref{fig:imageClassification}. While VGG11 and EfficientNet have the same average precision, they have a relatively low similarity score (\cref{fig:jaccard} B1 - .82). 

\subsubsection{Tversky Containment for Ensemble methods}
In machine learning, a common strategy for improving models is called ensemble learning: multiple model’s predictions are combined to produce a better outcome. To get the best results, models that make different predictions should be combined.  Using either a visualization such as SetMLVis or the Tversky similarity index, we can find models that make diverse predictions. 

The Tversky index provides insights into how well one model's predictions are contained within another's. This helps determine if Model B makes any predictions that Model A does not. For example, if Model A detects 100 objects and Model B detects 90 of the same objects then Model B is entirely contained in Model A (Tversky similarity index of 1). Combining these two models would be fruitless, as the results of the ensemble method would be the same as simply running Model B. By using the Tversky Index, we can effectively identify models with diverse predictions, ensuring that combining them in an ensemble leads to improved performance by capturing a broader range of patterns and anomalies.

\section{Conclusion}

Selecting and comparing machine learning models is a challenging task. Practitioners often rely on aggregate metrics that obscure performance differences or manually inspect instance-level predictions, which is time-consuming. Typical model comparison workflows evaluate models against the ground truth and use those evaluations to compare the models. We propose a new approach: directly comparing models against one another by organizing their predictions into sets. This method makes comparing models more intuitive and efficient by highlighting different model strengths and weaknesses through pertinent data subsets.

We generalize this approach by outlining four criteria that machine learning model outputs must meet to be placed into sets. The applicability of our approach is demonstrated through comparisons of clustering, regression, classification, and object detection models. These case studies reveal how set visualizations offer insights beyond conventional comparison techniques.

Finally, we demonstrate the effectiveness of using sets by developing an interactive visualization system to compare object detection models: SetMLVis. SetMLVis uses an UpSet-style visualization to connect aggregate set overviews to instance-level data in an easy-to-use Jupyter widget interface. We conduct a mixed-methods evaluation comparing SetMLVis to a comparable interface without set visualization, examining the impact sets have on task accuracy and cognitive workload for model comparison tasks. Our results indicate that using sets improves accuracy on model comparison tasks and reduces cognitive workload across all tasks. Future work could explore how sets and set visualizations can aid in model iterations and in creating new aggregate metrics to more holistically understand model comparisons.

\begin{acks}
Research was sponsored by the DEVCOM Analysis Center and was accomplished under contract numbers W911NF-22-2-0001 and W911QX-23-D002. The views and conclusions contained in this document are those of the authors and should not be interpreted as representing the official policies, either expressed or implied, of the Army Research Office or the U.S.\ Government. The U.S.\ Government is authorized to reproduce and distribute reprints for Government purposes notwithstanding any copyright notation herein.
\end{acks}

\bibliographystyle{ACM-Reference-Format}
\bibliography{setmlvis}


\end{document}